\documentclass[natbib]{svjour3}                     
\smartqed  
\usepackage{graphicx}
\newcommand{\aap}{{Astron. Astrophys.}}

\newcommand{\apjl}{{Astrophys. J.}}

\newcommand{\apjs}{ApJS}

\def\simgreat{\lower2pt\hbox{$\buildrel {\scriptstyle >}
   \over {\scriptstyle\sim}$}}
\def\simless{\lower2pt\hbox{$\buildrel {\scriptstyle <}
   \over {\scriptstyle\sim}$}}

\def\msun{\rm {M_\odot}}

\def\mt{m_{\rm \bullet, t}}
\def\mbhuno{m_{\bullet,1}}
\def\mbhdue{m_{\bullet,2}}
\def\kms{\rm {km\,s^{-1}}}
\def\mdisc{M_{\rm d}}
\def\sigmadisc{\Sigma_{\rm d}}
\def\rdisc{R_{\rm d}}

\begin{document}

\title{Massive binary black holes in galactic nuclei and their path to coalescence}

\author{Monica Colpi
}


\institute{Monica Colpi \at 
    Department of Physics G. Occhialini\\
             University of Milano Bicocca, Piazza della Scienza 3, I20123, Milano, Italy\\
             Istituto Nazionale di Fisica Nucleare (INFN) - Milano Bicocca, Piazza della Scienza 3, I20123, Milano, Italy\\
                              email:~{monica.colpi@unimib.it}           }

\date{25 June 2014, to appear in Space Science Reviews - DOI: 10.1007/s11214-014-0067-1}

\maketitle

\begin{abstract}
Massive binary black holes ($10^{5}\,\msun-10^9\,\msun$)
form at the centre of galaxies that experience a merger episode. They are expected to 
coalesce into a larger black hole, following the emission of gravitational waves.
Coalescing massive  binary black holes 
are among the loudest sources of gravitational waves in the Universe, and the detection of these events is at the frontier of contemporary astrophysics.
Understanding the black hole binary formation path and dynamics in galaxy's mergers is therefore mandatory. A key question poses:
during a merger, will the black holes descend over time on closer orbits, form a Keplerian binary and coalesce shortly after?
Here we review progress on the fate of black holes in both major and minor mergers of galaxies, either gas-free or gas-rich, in 
smooth and clumpy circum-nuclear discs after a galactic merger, and in circum-binary discs present on the smallest scales inside 
the relic nucleus.

\end{abstract}


\section{Massive binary black holes as tracers of black hole seed formation and galaxy assembly, along cosmic history}

In the universe, {\it black holes} come in {\it two flavours}: the "stellar black holes" relic of the most massive stars, weighing  $\sim 5-30\,\msun$ \citep{ozel10}, 
and the "super-massive black holes" residing in the nuclear regions of galaxies which carry large masses, typically in excess
of  $10^8\,\msun$ \citep{vestergaard08}.
 The black holes of stellar origin are
observed in X-ray binaries as accreting objects, while the super-massive ones power 
the bright QSOs and the less luminous active galactic nuclei (AGN) \citep{merloni13}.  Super-massive black holes
are also observed in their quiescent state as massive dark objects in nearby galaxy spheroids \citep{gultekin09,kormendyho13}, and 
a compelling case is that of the Milky Way housing an (almost) inactive black hole of $4\times 10^6\,\msun$ \citep{ghez08,gillessen09}.
 
A black hole {\it desert} exists between $\sim  30\,\msun$ and $\sim 10^6\,\msun$.  These black holes are often called 
{\it intermediate mass} or {\it middleweight} black holes with boundaries of the desert zone that are not physically constrained, due to the lack of observations.
The {\it maximum} mass of a black hole of stellar origin can be as large as a few $\times 10^2\,\msun$, 
according to theoretical studies, depending on
the metallicity of the collapsing progenitor stars and on the role of radiative feed-back in limiting the final mass \citep{omukai01,heger03}. The {\it minimum} mass of super-massive black holes is not constrained as unknown is the process of formation
of these black holes \citep{volonteri10,ferrarapalla13}. 

Limits on the density of the X-ray background light (resulting from
accretion of an unresolved population of massive black holes) and on the local black hole mass density \citep{marconi04,merloni13}, suggest that super-massive black holes have grown their mass across cosmic ages 
through repeated episodes of accretion and via coalescences driven by galaxy mergers. This  has  led to 
the concept of black hole {\it seed} and black hole growth from seeds in concordance with the rise of cosmic structure. 
The characteristic  mass or mass interval of the
seed population is unknown and weakly constrained theoretically.  
Thus, aim of contemporary astrophysics is to disclose the mechanism of black hole seed formation 
through the detection of middleweight black holes in galaxies \citep{reines13}.

The discovery of tight correlations between the black hole mass $M_\bullet$ and
stellar velocity dispersion $\sigma_*$,  and between $M_\bullet$ and the stellar mass of the spheroid $M_*$ ($M_\bullet/M_*\sim 10^{-3}$)
highlighted the existence of a process of symbiotic evolution between black holes and galaxies \citep{marconihunt03,haring04,ferrareseford05,gultekin09,kormendyho13}.
The current interpretation is that the huge power emitted by the central black hole when active may have affected the rate of star formation in the host, turing  the galaxy into a red and dead
elliptical \citep{mihos96,dimatteo05,hopkins06}. At present, there is a live debate on whether the correlation extends 
to bulge-less disc galaxies or in general to lower mass galaxies \citep{kormendyho13}.  Lower mass galaxies are expected to house lighter super-massive black holes, according
to the above relations. Thus low-mass galaxies are the preferred sites for the search of the middleweight black holes in the desert zone.
Many galaxies (up to 75\%) host at their centre a Nuclear Star Cluster, a compact sub-system of stars with mass $M_{\rm NSC}$
typically  $\simless 10^7\,\msun$, and half-mass radius of $\simless 10$ pc. In a number of Nuclear Star Clusters a central middleweight black hole has been discovered which co-habit the cluster \citep{ferrareseNSC06}.
Less tight correlations have been found between  $M_{\rm NSC}$ and the mass $M_*$ of the host galaxy \citep{scottgraham13}, indicating 
that while in bright spheroids the presence of a central black hole appears to be compulsory \footnote{See \cite{gerosa14} for
missing black holes in the brightest cluster galaxies following black hole coalescence and ejection by gravitational recoil.}, in less bright (disc) galaxies
a Nuclear Star Cluster, with or without a central black hole, is preferred. 

The search for middleweight black holes in less massive disc galaxies will be central for understanding 
the process of formation and co-evolution of black holes and galaxies \citep{reines13,kormendyho13}. 
But, this is not the only strategy. 
A powerful and promising  new route exists to unveil infant black holes, forming
at high redshift $z\simless 15$:  this is the route proposed by {\sl The Gravitational Universe},
the science theme selected by ESA for the next large mission L3, {\it for the search and detection of low frequency gravitational waves from
coalescing binary black holes in merging galactic halos} \citep{whitepaper13}.

\begin{figure}
\includegraphics[width=0.75\textwidth]{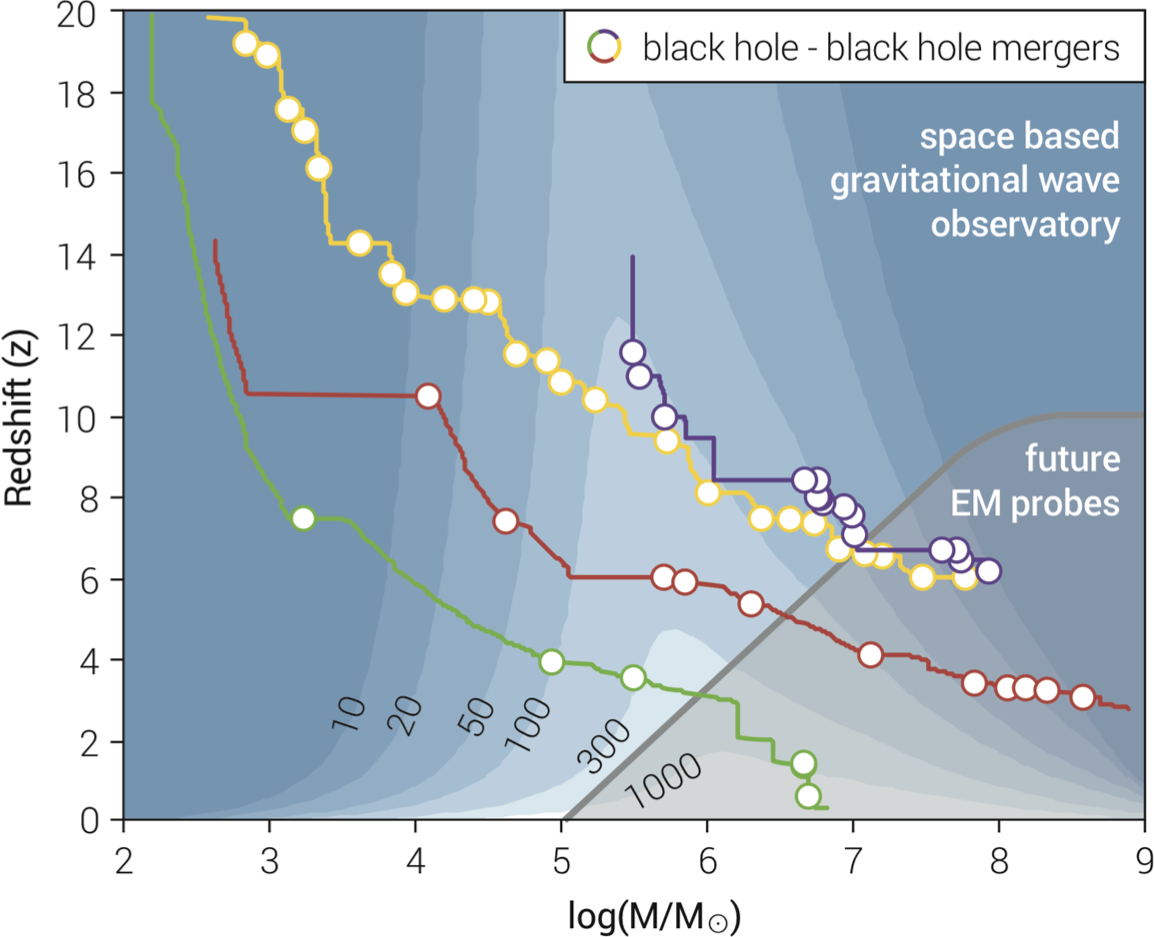}
\caption{
Paths of black holes forming at high redshift from light ($10^{2-3}\,\msun$) and heavy
($10^{5-6}\,\msun$) seeds.  The black holes evolve along tracks,  
in the mass versus redshift diagram, as they experience accretion episodes and coalescences with other black holes. Circles mark
the loci of  black hole coalescences.   Four paths are selected:
two ending with a black hole powering a $z \sim 6$ QSO (starting from a massive seed, blue curve, and from a seed resulting from 
the collapse of a massive metal-free star, yellow curve); 
a third ending with a typical $10^9\,\msun$  black hole in a giant elliptical galaxy (red curve); and finally 
the forth ending with the formation of a Milky Way-like black hole (green curve). The tracks are obtained using state-of-the-art semi-analytical merger tree models. The grey transparent area in the bottom right corner roughly identifies the parameter space accessible by future electromagnetic probes which will observe black holes powered by accretion. Over-lied are 
contour levels of constant sky and polarisation angle-averaged Signal-to-Noise-Ratios (SNRs) for eLISA, for equal mass non-spinning binaries as a function of their total rest frame mass \citep{whitepaper13}. It is remarkable that black hole mergers can be detected by eLISA with a very high SNR across all cosmic ages. Courtesy of \cite{whitepaper13}.}
\label{fig:bhmasszplot}
\end{figure}

According to the current paradigm of the $\Lambda$-CDM cosmology, galaxies form
following the baryonic infall of gas into collapsing dark matter halos and assemble hierarchically through mergers
of (sub-)galactic units \citep{whiterees78}. Black hole seeds growing in these pristine merging halos inescapably undergo coalescences \citep{volonteri03}. This is illustrated in
Figure~\ref{fig:bhmasszplot}, where we show characteristic tracks of black holes along cosmic history computed from
semi-analytical models of galaxy formation \citep{volonterinatarjan09}. The tracks are reported  in the mass versus redshift plane, from $10^2\,\msun$ up to
$10^9\,\msun$ and for $0\simless z\simless 20$: black holes
form from seeds of different masses (according to different seed formation models) and grow via accretion and mergers, the last denoted as
circles in the diagram.  The gravitational wave signal from these mergers will be detected with a high signal-to-noise ratio from 
the forthcoming science mission of {\sl The Gravitational Universe}, eLISA \citep{whitepaper13}. This will allow to explore black hole seed formation and evolution as early as $z\simgreat 10$, just at the end of the dark ages but well before the epoch of cosmic re-ionisation of the intergalactic hydrogen. 

Black holes come in binaries when two galaxies merge, and the gravitational wave signal emitted at coalescence 
offer a unique environment to measure, with exquisite precision,  
the black hole masses and spins, every time there is a merger \citep{gwnotes13}.  Thus, it has become mandatory to study how
and when binary black holes form and evolve inside galactic halos, during the formation of cosmic structures. This is a multi-face problem
crossing the boundaries between astrophysics and cosmology. 

This review aims at describing one aspect of the astrophysics of binary black holes: that of their dynamics in merging galaxies. 
This is a central problem if we want to 
consider binary black holes as powerful sources of gravitational waves and as unique tracers of the cosmic assembly of 
galaxies, as proposed in {\sl The Gravitational Universe} \citep{whitepaper13}. This problem carries some resemblance to 
the problem of formation, in stellar dynamics, of close binary neutron stars fated to coalesce:
rapid shrinking of the star's orbits occurs through phases of unstable mass transfer and common envelope evolution to avoid 
the risk of supernova disruption,
before the two compact stars reach the phase of inspiral by gravitational waves. Likewise, binary black holes experience
a phase of rapid sinking by dynamical friction when the two galaxies merge, but only after experiencing 
close encounters with stars and strong coupling with gas, they enter the phase of gravitational wave driven inspiral.
In the next sections, we will describe the fate of massive black holes in merging galaxies, indicating whether they form
close binaries ready to coalesce or wide pairs fated to wander in the host galaxy, and the conditions for this to happen. 

Section 2  starts with a brief historical recollection of the problem of black hole dynamics in stellar environments, and expands
these early findings to account for recent advances in this field. Section 3 addresses the problem of black hole dynamics
in  gas-rich galaxy major and minor mergers, while Section 4 describes the fate of black holes in gaseous circum-nuclear and circum-binary gas discs.
Section 5 presents a short summary of the timescales along the path to coalescence, and 
Section 6 contains the main conclusion and the directions in which the field may evolve into.

\section{Binary black holes in stellar environments}
\label{sec:intro}
\subsection{Super-massive black hole binaries in galactic nuclei}
\label{sec:low-redshift}
In a pioneering Letter to {\it Nature}, Begelman, Blandford and Rees (1980) write: "There are straightforward reasons for surmising that
super-massive black hole binaries exist: mergers between galaxies appear to be frequent;  cD galaxies in clusters or groups quite probably formed in this manner; and there is direct evidence that the near-by active galaxy Centaurus A is a merger product."  In particular, they correlate the bending and apparent precession of radio jets, observed in a number of active nuclei, with the presence of two black holes
in a binary, exploring 
the dynamics of its formation inside the violently relaxed {\it stellar core} of the newly formed galaxy.

Begelman et al.  depicted the existence of three main phases along the path to coalescence:
an early phase of {\it pairing} (phase I) under dynamical friction in the stellar bulge of the post-merger galaxy, ending 
with the formation of a close {\it Keplerian binary}; a phase of {\it hardening}  (phase II) during which the binary separation decreases 
due to energy loss by close encounters with single stars plunging on nearly radial orbits on the binary; a phase of  {\it gravitational wave inspiral} (phase III), ending  with the coalescence of the two black holes due to the emission of gravitational waves.
In all the phases, gravitational torques act to decrease the orbital angular momentum 
and energy of the black holes, 
to promote their pairing and sinking toward more bound states.

The black hole that forms has a new mass, new spin according to mass-energy conservation \citep{rezzolla08} and because gravitational waves carry away linear momentum, the new
black hole receives a gravitational recoil that can be as large as $\simless 5000\,\kms$ depending on
the orientation and magnitude of the black hole spins and orbital angular momentum at the time
of binary coalescence \citep {lousto13}.  Thus, an additional phase IV subsequent to merging should be considered corresponding 
to a {\it recoiling} black hole moving-inside or escaping-from it host galaxy \citep{gualandris08,merritt09,devecchi09}. This phase and the relation between spin, mass ratio and recoil are 
not considered here \citep{bogdanovic07,dotti10,centrella10} nor the observability of active black holes along their path to coalescence [we defer the reader to e.g. 
\cite{komossa06,schnittman11,bode10,eracleous11,liu11,komossa12,koss12,liu13,decarli13,comerford14,lusso14,fukun14}]. 

Returning to phase I,  it is known that 
dynamical friction against the stars acts on each black hole individually to cause their   
progressive sinking \citep{chandra43,bbr80,colpi99,yu02}, until they come close enough to form a {\it Keplerian binary}. 
As dynamical friction is proportional to the background density of stars and to the square of the black hole mass, more
massive black holes in denser environments sink more rapidly. 
In a stellar background of $N$ stars described by a singular isothermal sphere, with density profile $\rho_*=\sigma^2_*/(2\pi G r^2)$
and one-dimensional (1D) velocity dispersion $\sigma_*$, a black hole of mass $m_\bullet$ at distance $r$ sinks by dynamical friction on a timescale 

\begin{equation}
\tau_{\rm df}\sim  2\times 10^8 \ln^{-1} N\, \left ( {10^6\,\msun\over m_\bullet}\right )\left ({r\over 100 \,\rm {pc}}
\right )^2 \left ({\sigma_*\over 100 \,\kms}\right )\,\,\rm {yr}.
\label{dynfric}
\end{equation}
This timescale decreases with decreasing distance from the galaxy's nucleus, so that dynamical friction
becomes more and more rapid with orbital decay.  Eventually,  the black holes end forming a Keplerian system.
Binary formation occurs approximately when 
 the mass in stars enclosed in their orbit drops below twice the total mass of the binary $\mt=\mbhuno+\mbhdue;$
hereon $\mbhuno$ ($\mbhdue$) is the mass of the primary (secondary) black hole,
and $q=\mbhdue/\mbhuno\leq 1$ the mass ratio.  In a singular isothermal sphere, a Keplerian binary forms when $a^*_{\rm binary}\simeq G\mt/\sigma_*^2,$ i.e. at a separation comparable  to the gravitational sphere of influence of the black holes viewed as a single point mass $\mt$. 
Dynamical friction guides the inspiral, with no significant amplification of the eccentricity \citep{colpi99}, approximately down 
to $a^*_{\rm binary}$.
The weakening of dynamical friction is due to the high velocity that the black holes acquire when the form a binary, as the drag is 
inversely proportional to the square of the orbital velocity.

Phase I ends when the binary separation $a$ has decayed below 
\begin{equation}
a^*_{\rm hard}=a^*_{\rm binary}{\mu\over 3 \mt}\sim
{G\mu\over 3\sigma_*^2}\sim  0.1{q \over (1+q)^2}\left ( {\mt\over 10^6\,\msun}\right )
\left (\,{100 \, \break \kms\over \sigma_*}\right )^2\,\,\rm{ pc}, 
\label{a-hard-star}
\end {equation}
 where $\mu=\mt q/(1+q)^2$ is the reduced mass of the binary \citep{quinlan96,yu02,merritt05}. 
The hardening radius $a_{\rm binary}^*$ is  defined as the binary separation at which the kinetic energy per unit mass of the binary equals the kinetic energy per unit mass of the stars in the galactic potential.
During the hardening phase II, the black hole orbital energy and angular momentum are extracted via scattering of single stars off the binary, in close three-body encounters. 
As a single star impinging on the binary causes a fractional energy change of the order of $\sim  \xi m_*/\mt$ (where $\xi\sim0.2-1$ is a coefficient 
calculated after averaging over many star-binary scattering experiments), a large number of stars, or the order of 
$\sim \mt/ m_*$, is necessary for a sizeable change of the binary binding energy  $E_{\bullet}=
G \mbhuno \mbhdue /2a.$
 The binary offers a cross section $A\sim \pi a G \mt/\sigma_*^2$ to the incoming flow of stars and this  leads to a hardening rate 
$s\equiv d(1/a)/dt\sim \xi \pi G\rho_*/\sigma_*$ for the semi-major axis $a,$ and a corresponding hardening time (independent on
the number $N$ of stars in the galaxy)
\begin{equation}
 \tau^*_{\rm hard} \sim {\sigma_*\over  \pi G \rho_* a}
 \sim 70 \left ({\sigma_*\over 100 \,\kms}\right) 
\left ({10^4\,\msun\,{\rm pc}^{-3}\over \rho_*}\right )
\left ({10^{-3}\,{\rm pc} \over a}\right ) \,\,{\rm Myr}.
\label{time-hard}
\end{equation}
Opposite to $\tau_{\rm df}$, the hardening time $\tau^*_{\rm hard}$ increases with decreasing $a$, as the binary cross section decreases with $a$. 
Thus, a potential stalling of the binary can occur at the smallest binary separations, during phase II. 

Phase III starts when the coalescence time driven by gravitational wave emission
\begin{equation}
\tau_{\rm gw}
\sim 5.4\times 10^8 f(e)^{-1}{(1+q)^2\over q }{a^4\over  \mt^3} \left ({1\over 0.001\,\rm{pc}}\right)^4 \left ({
10^6\,\msun\over \mt}\right )^3\,\, \rm{yr}
\label{time-gw}
\end{equation}
drops below $\tau^*_{\rm hard}$, where $f(e)=[1+(73/24)e^2+(37/96)e^4](1-e^2)^{-7/2}.$
The crossing condition, $\tau^*_{\rm hard}=\tau_{\rm gw}$ thus provides the binary separation at which
the black holes transits from phase II into III:
\begin{equation}
a^*_{\rm {II\to III}}=\left ({G^2\over c^5}{256\over 5\pi}\right )^{1/5}
\left ({\sigma_*\over \rho_*}\right )^{1/5}f^{1/5} (e)
\left ({q\over (1+q)^2}\right )^{1/5}\mt^{3/5}.
\label{star-I-II}
\end{equation}
If $\tau_{\rm gw}$ evaluated at $a^*_{\rm {II\to III}}$ exceeds the age of the universe, than the binary {\it stalls} and does
not reach coalescence. 
From equation [\ref{time-gw}], we can define as $a_{\rm gw}$ the distance at which
the coalescence time 
$\tau_{\rm gw}$ equals  the Hubble time
 $\tau_{\rm Hubble}$: 
\begin{equation}
a_{\rm gw}= 2\times 10^{-3} f(e)^{1/4}{q^{1/4}\over(1+q)^{1/2}}\left ({\mt \over 10^6\,\msun}\right )^{3/4}\left( {\tau_{\rm Hubble} \over 13.6\,{\rm Gyr}}\right)^{1/4}\,\,{\rm pc}.
\label{agw}
\end{equation}
Expressed  in units of the Schwartzschild radius $r_{\rm S}=2G\mt/c^2$
associated to $\mt$,  $a_{\rm gw}=1.4\times 10^4 (\mt/10^6\,\msun)^{-1/4} r_{\rm S}$ for the case of an equal mass circular binary.
Coalescence occurs as long as $a^*_{\rm {II\to III}}<a_{\rm gw}.$

According to equation~[\ref{time-hard}], for a wide interval of stellar densities and velocity dispersions, the coalescence time $\tau_{\rm gw}$, evaluated at $a^*_{\rm {II\to III}}$, 
is less than the Hubble time $\tau_{\rm Hubble},$ 
so that the binary is excepted to enter the gravitational wave driven regime shortly after it has become hard.
However, 
the estimate of 
$\tau^*_{\rm hard}$, in  equation~[\ref{time-hard}], severely {\it underestimates the true hardening time} since
a large number of stars in "loss cone" orbits is necessary to 
drive the binary down to phase III.
The loss cone in the black hole binary system is identified as the domain, in phase-space, populated by stars
with sufficiently low angular momentum, $J^2\simless J^2_{\rm lc}\simless 2G\mt a$, to interact with the binary. If hardening occurs at a constant rate $s$, 
the number of stars necessary to complete the hardening phase is as large as $N^{\rm lc}\sim (\mu/m_*)\ln(a^*_{\rm hard}/a_{\rm gw}),$ 
comparable to the mass of the binary.
In the case of massive black holes ($\mt>10^8\,\msun$)
in elliptical galaxies and spheroids, such a large reservoir of stars may {\it not} be available \citep{merrittlosscone13}.  

At the end of phase I, when stellar encounters begin to  
control the contraction of the newly formed binary, the black holes start ejecting stars from the loss cone at a high clearing rate. 
Refilling of stars in the phase-space 
requires a lapse time comparable to the two-body relaxation timescale $\tau_{\rm rel}\propto N$
which in galactic nuclei, viewed as spherical systems, is often longer than the Hubble time \citep{yu02}.
Thus, the lack of stars in phase-space causes the binary to {\it stall}, at a separation $a^*_{\rm stall}$ typically of $\sim 0.1-1$ pc,  much larger than $a_{\rm gw}$ (eq.~[\ref{agw}]). Thus the binary can not reach coalescence in a Hubble time, and this is referred to as {\it the last parsec problem}.  This represents an obstacle to the path to coalescence during the transit across  phases II and III,
for a large range of black hole masses, and mass ratios $q$  \citep{yu02}. We are therefore left with a major uncertainty on the estimate of
the {\it true hardening time} $\tau_{\rm Hard}$ which is expected to be closer to $\tau_{\rm rel},$ in the case of empty loss cone,
and to 
$\tau_{\rm hard}^*$ as given by equation~[\ref{time-hard}], in the case of full loss cone: thus,  $\tau_{\rm hard}^*<\tau_{\rm Hard}<\tau_{\rm rel}$.

The binary is a source of kinetic energy as it deposits in the stellar bath an energy 
 \begin{equation}
\Delta E_\bullet\sim E_\bullet(a_{\rm gw})\sim 
2\times 10^{55} f(e)^{-1/4}{q^{3/4}\over(1+q)^{3/2}}\left ({\mt \over 10^6\,\msun}\right )^{5/4}\left( {\tau_{\rm Hubble} \over 13.6\,{\rm Gyr}}\right)^{-1/4}\,\,{\rm erg},
\label{energy}
\end{equation}
in order to enter phase III. Compared to the binding energy of  a stellar bulge of mass $M_*$, $(3/2)M_*\sigma_*^2$, energy deposition accounts $\sim$ 10\%  of the total energy of the system, if one
assumes $\mt\sim 10^{-3} M_*$, a value of $\sigma_*\sim 100\,\kms$, and an equal mass binary of  $10^6\,\msun$.  Binary energy deposition
via encounters with single stars can create a stellar core in an otherwise steep density profile, due to star's ejection.
Stellar scouring has been observed in a number of core, missing-light elliptical galaxies that are at present indirect candidates
for black hole mergers \citep{milos01,kormendyho13,merrittbook13}. 
The binary carries a larger angular momentum at $a^*_{\rm hard}$ compared to the angular momentum at the onset of gravitational wave inspiral,
\begin{equation}
{J_{\bullet}(a^*_{\rm hard})\over
J_{\bullet}(a_{\rm gw})}
\sim 10 \,(1-e^2)^{7/16}\,{q^{3/8}\over (1+q)^{3/4}}\,\left ({\mt\over 
10^6\, M_\odot}\right )^{1/8}\,{100\, {\rm km\,s^{-1}}\over \sigma_*}\left( {\tau_{\rm Hubble} \over 13.6\,{\rm Gyr}}\right)^{-1/8}.
\end{equation}
From the equation it is clear that the binary can reduce $J_\bullet$ when transiting from phase II to III, increasing the eccentricity
during orbital decay.

After Begelman et al., the last parsec problem has been considered a major
bottleneck to the path of binary coalescence, and has motivated many studies \citep{milos01,yu02,merritt05}. 
Direct $N$-Body simulations of binary inspiral in isotropic, spherical galaxy models confirmed, on solid 
grounds, the stalling of the binary: the binary hardening rate $s$ was found to be proportional to the rate of repopulation
of loss cone orbits which in turn depends on $N$.  Simulations with a lower number of particles $N$ (corresponding to
shorter two-body relaxation times $\tau_{\rm rel}$) show rapid binary decay. By contrast, more realistic simulations with larger $N$ (longer $\tau_{\rm rel}$) display a much lower hardening rate $s$ for the binary \citep{preto11}.
The extrapolation of the result to the limit of very large $N$, as in elliptical galaxies or bulges of spirals, 
leads to stalling of the massive binary over a Hubble time.

\cite{yu02} noticed that if one drops the assumption of sphericity, the hardening time $\tau_{\rm Hard}$ is lower and can be less than 
the Hubble time in the case of less massive (power-law) galaxies which have a shorter $\tau_{\rm rel}$.  Spherical galaxies have all stars on centrophobic  orbits, whereas 
galaxies with a high degree of axisymmetry and triaxialily host a significant fraction of stars on centrophilic  orbits, such box orbits, which pass arbitrarily close to the binary and have low angular momentum. Furthermore, chaotic orbits in  steep
triaxial potentials can enhances the mass flux into the loss cone region \citep{merrittpoon04}.
Some non axisymmetric potential can also excite bar instabilities causing a flow of stars toward the binary \citep{berczik06}. 

 Binary stalling has been recently challenged in models of galaxy's mergers.
 A number of direct $N$-Body simulations indicate that the end-product of a merger is not a spherical galaxy  \citep{berczik06,khanjustmerritt11,preto11,khan13,wang14}. 
 The  new galaxy retains substantial amount of rotation or/and a large degree of asphericity or triaxiality such that
  the binary is seen to harden at a rate independent of $N$, as if the loss cone were fully refilled, or as if an
$ N$-independent mechanism (collisionless relaxation) provides a supply of stars in loss cone orbits.
In light of these findings the last parsec problem appears today as an artefact of the oversimplifying assumption
 of sphericity of the relic galaxy, and
that more realistic models, simulated starting from {\it ab initio} conditions,  point in the direction of hardening times ranging between 0.1
and a few  Gyrs, for the models explored. \footnote{This view, however,
has been criticised by \cite{vasiliev13} who compared the evolution
of binary black holes in spherical, axisymmetric and triaxial equilibrium galaxy models. 
They find that the rate of binary hardening exhibits a significant $N$-dependence in all the models, in the investigated range of $10^5\leq N\leq 10^6$. Their hardening rates are substantially lower than those expected if the binary loss cone remained full, with 
rates between the spherical and non-spherical models differing in less than a factor of two.  This finding seems to cast doubt on claims that 
triaxiality or axisymmetry alone are capable of solving the final-parsec problem. 
Vasiliev and co-authors invite caution in extrapolating results to galaxies with high values of $N$ 
until all discrepancies or intrinsic differences between equilibrium models and merged galaxy models are not understood deeply. }
An interesting corollary of these investigations is that in non-spherical models the binary eccentricity $e$ is seen to increase to 
 values very close to 1 \citep{preto11,khanjustmerritt11}, indicating rapid 
 transfer of angular momentum to stars from the cumulative action of  many scatterings {\citep{sesanaecc10,dosede12}.

As final remark, alternative mechanisms exist that can cause the contraction of the binary orbit, such as recycling of  stars
ejected by the binary on returning eccentric orbits \citep{milos03returning}, 
massive perturbers scattering stars into loss cone orbits \citep{perets07}, or a third black hole in a trio encounter.
In the latter case, the third closely interacting black hole can harden the binary due to eccentricity oscillations \citep{blaes02}, 
or  energy exchange in the three-body scattering, or can repopulate the loss cone having perturbed the underlying gravitational potential
of the host galaxy \citep{hoffman07,kulkarni12}.

So far, we considered  the hardening of massive black hole binaries in massive galaxies. But, a question to investigate 
is related to the evolution of middleweight black holes of $\sim 10^{3-4}\,\msun$ which tend to inhabit smaller mass halos with shorter relaxation timescales, and
that form at high redshift when the universe was younger. This narrows down the interval of time accessible for
hardening.  Is there a last parsec problem?
Extrapolating the results to middleweight black hole masses may not be straightforward, and 
in the next subsection we shortly explore the hardening in this regime.

\subsection{Middleweight binary black holes in stellar environments: hardening or stalling in high redshift nuclei?}
\label{sec:high-redshift}
Studying the hardening and coalescence of middleweight black hole binaries with  $\mt \simgreat 10^4\,\msun$  
is of  importance as black holes of this mass are primary sources for  eLISA, as illustrated in 
Figure~\ref{fig:bhmasszplot}.
In the figure, black hole coalescences occur at a rate equal to the rate of merging of their parent dark matter halos controlled
by dynamical friction only.  The underlying assumption is there is no or negligible delay 
between the merger of the halo and that of the nested black holes, caused by the potential stalling of the binary.
At present, whether black hole binaries at very high redshift are able to reach coalescence in the short 
cosmological time lapse between black hole seed formation and halo mergers is unclear (paper in preparation).

As an exercise and for illustrative purposes, one can compute limits upon the density $\rho_*$ and velocity dispersion $\sigma_*$ that 
a massive stellar cluster should have to allow rapid hardening of the binary during phase II.  In star clusters the density
and velocity dispersion are functions of distance. Thus, one should consider $\rho^*$ and $\sigma_*$ are 
characteristic values of the central region
of a massive star cluster in an hypothetical galactic nucleus.

Focus on the case of a black hole forming at $z_{\rm form}$ (e.g. $\sim 20$) and coalescing with another black hole, following a halo-halo merger
at  $z_{\rm coal}$ ($\sim 15$), or the case of two adjacent mergers between redshift $z_1$ and $z_2$ over a short interval of cosmic time.  
The time lapse  $\Delta \tau_{\rm lapse}$ 
 can be as short as $\simless 0.1$ Gyr, or more conservatively as short as 1 Gyr.
  In Figure~\ref{fig:high-low-redshift} we plot, in the $\sigma_*-\rho_*$  plane, the lines of constant  $\tau_{\rm rel}=0.34 \sigma^3_*/(G^2 m_* \rho_*\ln\Lambda)$ (with $m_*=1\,\msun$ and $\ln\Lambda\sim 10$)
corresponding to a cosmic time lapse $\Delta \tau_{\rm lapse}$ equal to 0.1 Gyr and 1Gyr, respectively.
The solid lines in Figure~\ref{fig:high-low-redshift} refer to the loci where $\tau_{\rm rel}=\Delta \tau_{\rm lapse}$, so that characteristic densities higher than
\begin{equation}
\left ( {\rho_*\over 1.6\times 10^7\,\msun {\rm pc}^{-3} }\right )_{\rm rel} \simgreat \left ({\sigma_*\over 100\,\kms}\right )^3 
\left ({0.1\, {\rm Gyr}\over \Delta \tau_{\rm lapse}}\right )
\label{rho-taurel}
\end{equation}
are requested, at a fixed $\sigma_*$,  to allow for binary hardening on the relaxation timescale (corresponding to the empty loss cone regime).
\begin{figure}
  \includegraphics[width=0.80\textwidth]{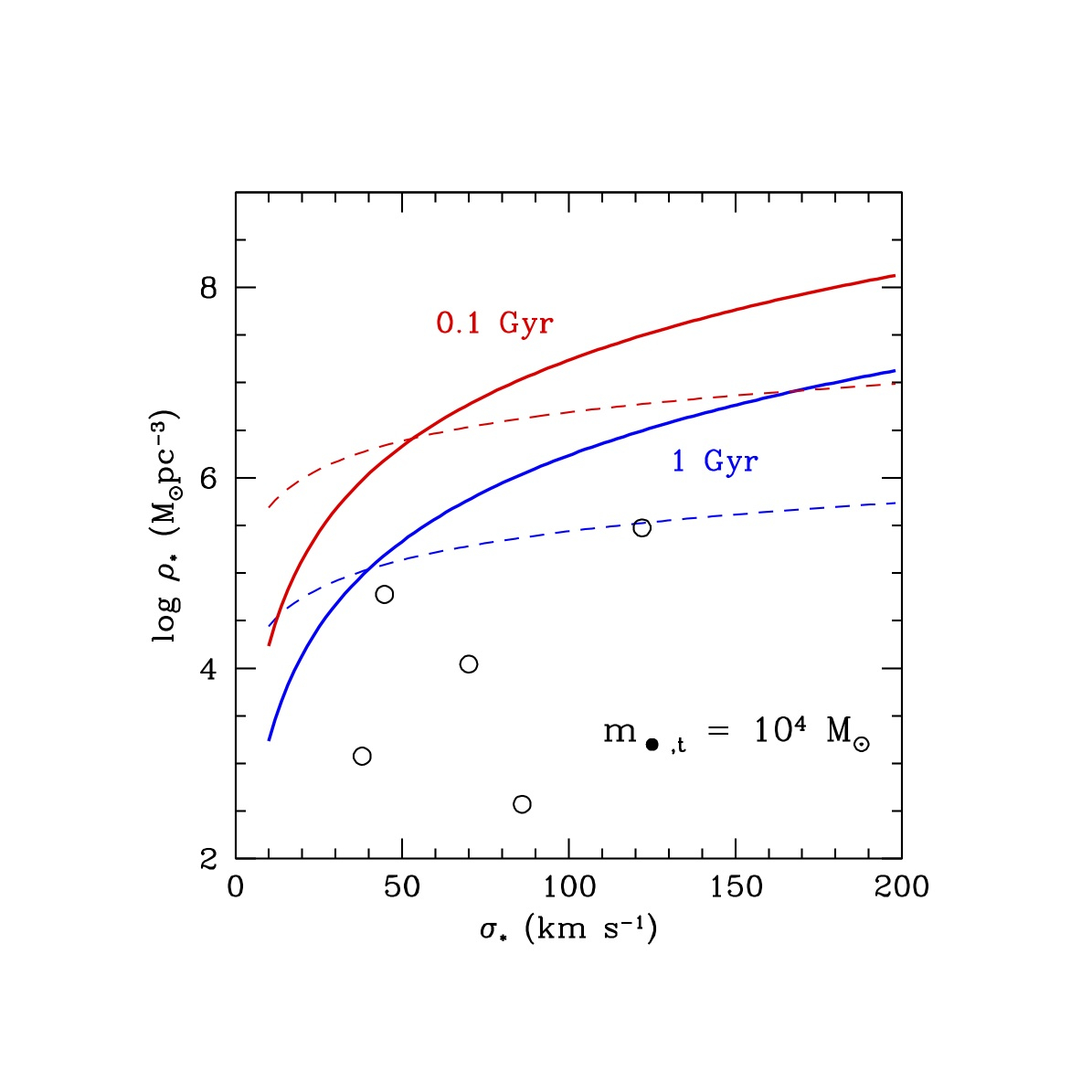}
\caption{Stellar mass density (in units of $\msun\,{\rm pc}^{-3}$) versus velocity dispersion (in units of $\kms$) of hypothetical 
nuclear star clusters hosting middleweight black hole binaries 
on their path to coalescence which harden via single-binary encounters with solar mass stars:  solid lines refer to the loci in the ($\sigma_*,\rho_*$) plane 
given by eq.~\ref{rho-taurel} where the central relaxation time
$\tau_{\rm rel}$ equals $\Delta \tau_{\rm lapse}$.
The upper-red (lower blue) solid line refers to $\tau_{\rm rel}=\Delta \tau_{\rm lapse}=0.1$ Gyr (1 Gyr).
Dashed lines refer to an equal-mass black hole binary of $10^4\,\msun$ and refer to the loci where the hardening time $\tau_{\rm hard}^*$, given by eq.~[\ref{time-hard}], equals
 $\Delta \tau_{\rm lapse}$. 
The hardening time is computed at a black hole binary separation $a_{\rm gw}$ given by eq.~[\ref{agw}].
The upper (lower) dashed line refers to $\tau^*_{\rm hard}=\Delta \tau_{\rm lapse}=0.1$ Gyr (1 Gyr).
Empty circles refer to mean
stellar densities and velocity dispersion (calculated using the virial theorem) for five Nuclear Star Clusters of known mass and half-mass radii
\citep{seth08,merrittlosscone13}.
}
\label{fig:high-low-redshift}
\end{figure}
The dashed lines in Figure~\ref{fig:high-low-redshift} refer instead to the loci where  $\tau^*_{\rm hard}(a_{\rm gw})=\Delta \tau_{\rm lapse}$,
as given by equation~[\ref{time-hard}] (corresponding to the full loss cone regime), for a black hole binary of $\mt=10^{4}\,\msun$ (upper dashed curve  for 
$\Delta \tau_{\rm lapse}$ equal to 0.1 Gyr, lower dashed curve for 1 Gyr).
 This condition implies 
\begin{equation}
\left ( {\rho_*\over 6\times 10^6\,\msun {\rm pc}^{-3} }\right )_{\rm hard} \simgreat \left ({\sigma_*\over 100\,\kms}\right )  \left ( {10^4\,\msun \over \mt}\right )^{3/4}
\left ({0.1\, {\rm Gyr}\over \Delta \tau_{\rm lapse}}\right )^{5/4}. 
\label{rho-limit}
\end{equation}
Figure~\ref{fig:high-low-redshift} shows that true hardening, at early cosmic epochs, 
requires densities in excess of $\simgreat 10^{6-8}\,\msun \,{\rm pc}^{-3}$ and comparatively
low dispersion velocities $\simless 70\,\kms$ to meet the conditions for coalescence in the short time lapse $\Delta \tau_{\rm lapse}$ of 0.1 
or 1 Gyr. 
We do not know if such dense stellar environment were present at the centre of unstable pre-galactic discs. Today Nuclear Star Clusters, plausible candidates to harbour a central middleweight black hole, have all densities and velocity dispersions that do not meet
this condition. Non-equilibrium conditions and/or the presence of
gas, abundant in pre-galactic discs, may be instrumental in taxing the black holes to small separations, in this interval of masses, and
at earlier cosmic epochs. 
Thus, a key question to pose is whether {\it gas} can {\it fasten} the transition along the three phases of pairing, hardening  and gravitational wave driven inspiral and this question will be addressed in the incoming sections.

\section{Black holes dynamics in gas-rich mergers}
\label{sec:pairing}

Merging galaxies which are the sites of formation of binary black holes are expected to contain large concentrations of cold gas (unless
one considers mergers between today elliptical galaxies only). 
This inevitable abundance of gas, in particular in high redshift disc galaxies, motivated us to inquiry into 
the role of gas dynamics as an alternative in the process of black hole hardening and coalescence. 

In this section, we review 
the pairing of black holes in gas-rich merging galaxies following their dynamics {\it ab initio} to highlight the key role played by gas in affecting the black hole inspiral and the remarkable difference between {\it major} and {\it minor} mergers.  Major mergers refer to interactions between galaxies of
comparable mass, while minor mergers refer to interactions between a primary massive galaxy and a 
less massive galaxy, typically with mass ratio 1:10, and below.  
Boundaries  among major or minor mergers are not sharp, as in many cases, the various outcomes depend also on the internal structure
and gas content of the interacting galaxies.

\subsection {Major mergers and the formation of a Keplerian binary}
\label{sec:majormergers}

The study of black hole dynamics 
 in {\it gas-rich mergers} dates back to \citep{mayer07}, yet it is still in its infancy.
The rich physics involved and the high computational demand require state-of-the-art simulations, and the body of data 
is still inhomogeneous, fragmented and incomplete.  While
black hole dynamics in collisionless mergers of spherical galaxies has been explored (with direct $N$-Body codes) starting from 
galaxies on close elliptical bound orbits and followed mainly during the hardening phase \citep{khanjustmerritt11}, black hole dynamics in mergers between gas-rich disc galaxies has been 
studied starting from cosmologically motivated (parabolic) orbits, during the pairing phase over
separations $\simgreat 10$ kpc, down to the scale ($\simless 10$ pc) when the black holes
form a Keplerian binary \citep{mayer07,colpi09,colpidotti11,chapon13,mayer13}.  Further hardening has been later explored in dedicated simulations \citep{escala05,dotti06,dotti07,dotti09,fiacconi13}. 

Disc galaxies, as observed at low redshifts, are multi-component systems comprising  
a collisionless dark matter halo, a stellar disc which coexists with 
a multi-phase gaseous disc, and a central bulge housing (when present) a massive black hole. 
Simulating a collision between two disc galaxies with central black holes thus requires simulating
the dynamics of the collisionless components (dark matter and stars) jointly with that of
gas which is dissipative, and thus subject to cooling, star formation, shock heating and stellar feed-back. 

There are many simulations of disc galaxy mergers in the literature (e.g. \citep{hopkins13}), but there exists 
only a limited number in which the black hole dynamics is followed self-consistently from the  $\simgreat$ kpc scale 
down to scales $\simless 10$ pc.
When two galaxies merge, the two black holes are customarily assumed to merge promptly and form a single black hole. 
A recent set of $N$-Body/Smooth Particle Hydrodynamic simulations exists which follow the dynamics of black holes 
from the 100 kpc scale, typical of a merger, down to a scale 
$\simless 10$ pc, and which include star formation and feed-back \citep{vanwasse12,vanwasse14}.  
There exists a further class of SPH or/and Adaptive Mesh Refinment (AMR)  simulations which have enough resolution to witness the formation of a Keplerian binary 
on the $\sim 1$ pc scale, but which 
treat the gas thermodynamics via a phenomenological
energy equation, in the form of a polytrope \citep{mayer07,chapon13}.

Equal mass mergers are disruptive for both progenitor galaxies.
The galaxies first experience a few close fly-by during which tidal forces start to tear the galactic discs
apart, generating tidal tails and plumes.  The discs sink by dynamical friction against the dark matter background, and
the massive black holes follow passively the
dynamics of the bulge and disc they inhabit. Prior to merging, during the second pericentre  passage, strong spiral patterns appear in
both the stellar and gaseous discs: non axisymmetric torques
redistribute angular momentum so that as much as 60$\%$ of the gas
originally present in each disc of the parent galaxy is funnelled 
inside the inner few hundred parsecs of the individual galaxy centres. The black
holes, still in the pairing phase, are found to be surrounded by a
rotating stellar and gaseous disc.

Later, the gasoues discs eventually merge in {\it a single
massive rotationally supported nuclear disc} of
$\simless 100$ pc in size, now weighing $\sim 10^9\,\msun$.  The disc develops gravo-turbulence (with velocities $\sim 60-100 \,\kms$) that guarantees a Toomre parameter $Q\simgreat 2$. This prevents fragmentation of gas into stars on the timescale necessary for the black holes to form
a Keplerian binary (a few Myr).
This short sinking timescale
comes from the combination of two facts: that gas densities are higher
than stellar densities due to the dissipative nature of
the interaction, and that the black holes move relative to the
background with mild supersonic velocities. Under these conditions, the hydro-dynamical drag is the
highest  \citep{ostriker99}.  
The subsequent evolution is described in Section~\ref{sec:nucleardisc}.

During final revision of this review, a new dedicated $N$-Body/SPH simulation by \citep{roskar14} 
of two Milky-Way-like galaxy discs with moderate gas fractions, has been carried out at parsec-scale resolution,
 including a new model for radiative cooling and heating in a multi-phase medium, star formation and feedback from supernovae. 
The massive black holes weighing $\sim 10^6\,\msun$ are form a pair at a separation of $\sim 100$ pc which gradually
spirals inward. However, due to the strong starburst triggered by the merger, the gas in the centre most region is 
evacuated, requiring 
$\sim 10$ Myr for the nuclear disc to rebuild. The clumpy nature of the interstellar medium has a major impact on the dynamical evolution of the pair now subjected to stochastic torquing by both clouds and spiral modes in the disc.  These effects combine to delay the orbital decay of the two black holes, just in phase I of gas-dominated dynamical friction.  An inspiral timescale of 
$\sim 100$ Myr is found in this simulation which is smaller compared to that 
estimated in collisionless mergers, but longer of a factor at least 10 compared to the case of mergers with a single-phase gas. The result is in line with what found in \cite{fiacconi13} (see 
Section~\ref{sec:clumpydisc}) who describes black hole dynamics in clumpy nuclear discs. We notice however that a single run may not
suffice to pin down the  characteristic gas-dynamical friction timescale in dissipative mergers, and that the perturbations induced
by a population of massive clumps in the stellar component may alter the star's dynamics, prompting rapid refilling of the loss cone region around the two black holes, an effect that these simulations can not capture.

\subsection {Black hole paring in unequal-mass mergers} 
\label{sec:unequalmergers}
  
 \subsubsection{Collisionless unequal-mass mergers}
 \label{subsec:collisionlessunequal}

Early works on collisionless mergers of unequal-mass spherical dark matter halos \citep{governato94,taffoni03,boylan08} indicated that 
additional mechanisms are present, besides dynamical friction, 
that influence the structure and orbital evolution of the interacting halos (primarily the less massive, secondary): (i) progressive mass loss, or {\it tidal stripping}, induced by the tidal field of the main halo which reduces the mass of the secondary delaying the sinking by dynamical friction (the force scaling as the square of the satellite mass), and  (ii) {\it tidal heating}, i.e. the effect of 
short impulses imparted to bound particles 
within the secondary satellite galaxy by the rapidly varying tidal force of the primary which heats the system causing its (partial) dissolution \citep{taffoni03}.
Depending on the energy $E$ of the orbit  and its degree of circularity $\varepsilon$, on the relative mass concentration $c_{\rm s}/c_{\rm h}$
between satellite and main halo,
and on the initial mass ratio of the primary to the satellite halos  $M_{\rm h}/M_{\rm s}$,
the encounter can lead either to rapid merging toward the
centre of the primary halo (M), disruption (D), or survival (S) (when a
residual mass remains bound and maintains its identity,
orbiting in the main halo for a time longer than the Hubble time).
Figure~\ref{fig:taffoni} illustrates the various outcomes of these experiments. In this context, merging times can be described
by an empirical equation which accounts for the progressive mass loss of the secondary by tidal stripping and the 
progressive delay in the halo merging process 
\begin{equation}
{\tau_{\rm df,tidal}\over t_{\rm dyn}}\approx\, {\Theta (E,\varepsilon,c_{\rm s}/c_{\rm h})\over \ln(1+M_{\rm h}/M_{\rm s}(t))}\, {M_{\rm h}\over M_{\rm s}(t)}
\label{tau-merging}
\end{equation}
 where $\Theta$, function of the initial parameters,  and satellite mass $M_{\rm s}(t)$ are computed from the numerical simulation.
Figure~\ref{fig:taffoni} shows the fragility of less concentrated satellites  to
dispersal and disruption. 
These findings anticipate the possibility that unequal-mass mergers may release black holes on peripheral orbit inside the primary, due 
to tidal stripping of the less massive galaxy prior completion of the merger.

\begin{figure}
\hbox{
  \includegraphics[width=0.33\textwidth]{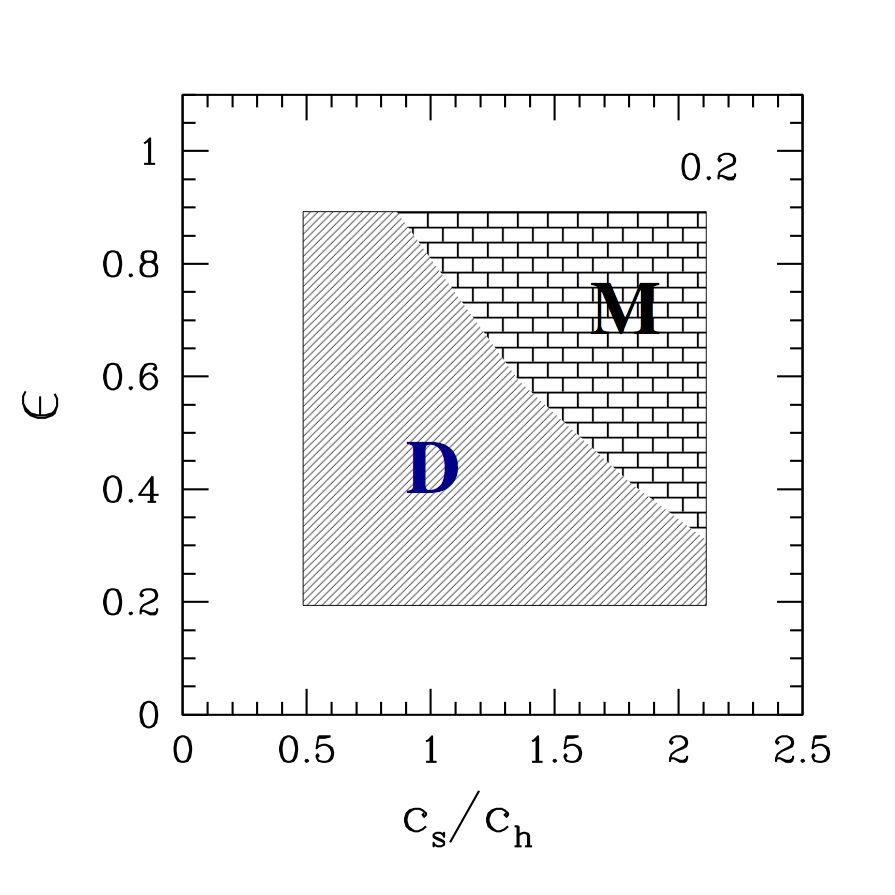}
  \includegraphics[width=0.33\textwidth]{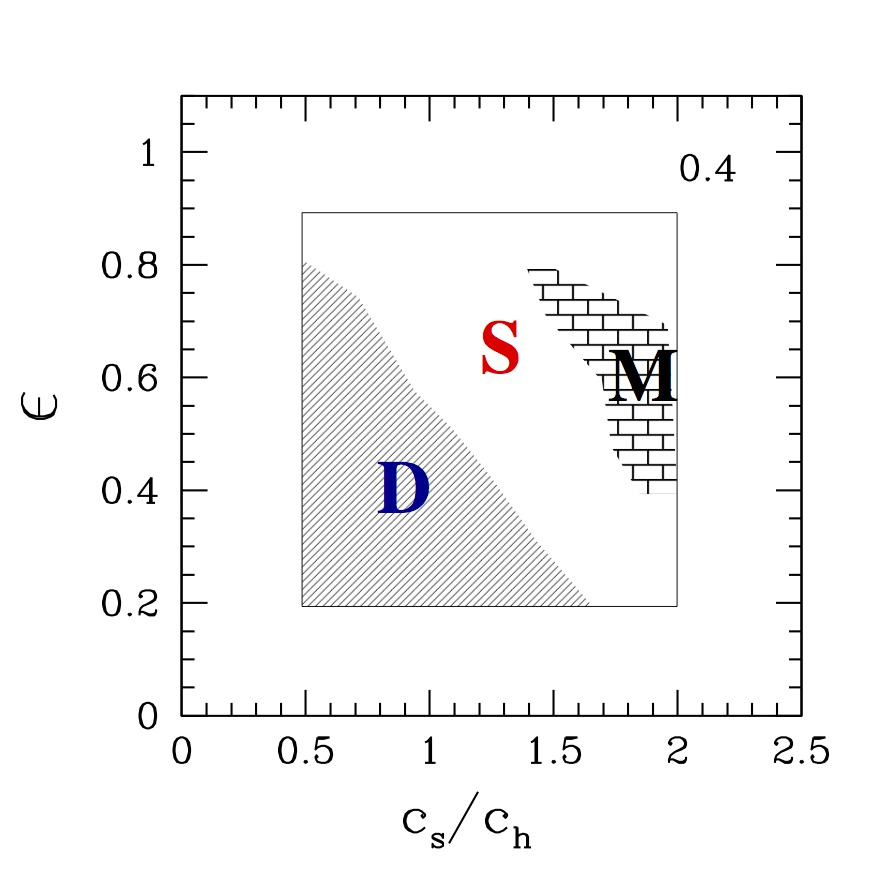}
   \includegraphics[width=0.33\textwidth]{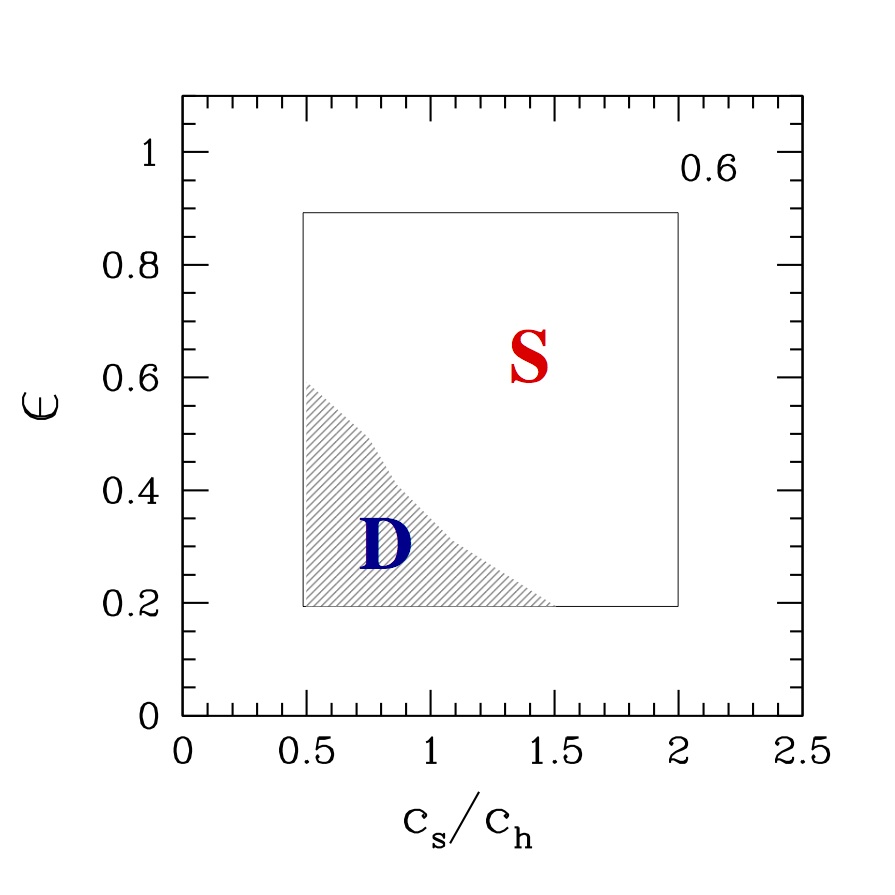}}
\caption{Outcome of simulations of unequal mass mergers between spherical (dark matter) halos described with a NFW profile \citep{navarro96}, and described 
in the plane circularity $\varepsilon$ versus relative concentration $c_{\rm s}/c_{\rm h}$ (measuring the ratio between the
scale radii of the two halos, defined as in \citep{navarro96,taffoni03}). 
The figure depicts the life diagram of a satellite halo with mass $M_{\rm s}/M_{\rm h}=0.01$. Each plot is labelled by the value $x(E)$, the radius of the circular orbit (in units of the half mass  of the main halo) at the onset of dynamical evolution. We
identify the regions corresponding to merger (M) of the satellite into the main host halo, disruption against the background (D), and survival (S) of part of the satellite in the periphery of the main halo.  Satellite halos with low concentration on less circular orbits are fragile to disruption, while satellite halos  on wide orbits and high concentration preserve their identity. Mergers are preferred in correspondence of high concentration and close circular orbits. Courtesy of  \cite{taffoni03}.}
\label{fig:taffoni}
\end{figure}

\subsubsection{Gas-rich unequal-mass mergers}
\label{subset:gasunequal}
Recent suites of $N$-Body/SPH  simulations of unequal-mass galaxy mergers 
have highlighted the occurrence of new key features in the dynamics of the discs and their embedded black holes that can be ascribed to differences in the central concentration of the interacting galaxies, and to the geometry of the encounter, but that go
beyond the results inferred in Section~\ref{subsec:collisionlessunequal}. These new simulations illustrate the pivotal role played by gas which acts, through its cooling, to enhance the
central mass concentration of the satellite and favours the sinking of the 
secondary black hole in the otherwise disrupted galaxy \citep{kaza05,callegari09,callegari11,vanwasse12,vanwasse14}.

In unequal-mass mergers, the secondary, less massive galaxy undergoes major 
transformations.   
In particular, if the merger is wet, i.e. if the gas fraction in the disc of the secondary is relatively high ($\simgreat 10\%$), tidal torques 
during the last peri-centre passage prior merging, trigger inflows which give rise to a nuclear starburst in the vicinity of the 
secondary black hole. This enhances the resilience of the galaxy's nucleus against tidal stripping due to the increased
stellar density and degree of compactness of the nuclear bulge, at the time 
the secondary starts interacting with the disc of the primary. The denser stellar cusp surrounding the secondary black hole
thus sinks rapidly toward the primary, dragging the black hole that reaches a separation of $\sim 100$ pc, close to the resolution
limit of the simulation. This is illustrated in Figure~\ref{fig:callegari1} for 1:4 and 1:10 wet  mergers, where the relative separation of the black holes is plotted against time
(heavy solid line). 
In Figure~\ref{fig:callegari1} we also plot the stellar and gas distribution at the end of the nuclear starburst that created a denser
stellar nucleus in the secondary.
The disc of the secondary, rather turbulent and clumpy due to star formation, is later disrupted by 
 ram pressure stripping by the gas of the primary. The secondary black hole continues its sinking toward the centre of
 the primary, being surrounded by the compact and massive star cluster. 
 In Figure~\ref{fig:callegari1}, we also contrast the results from dry, i.e. gas free mergers. In the absence of
the central starburst,  dry mergers leave the secondary black hole
wandering on a peripheral orbit at $\sim 1$ kpc away from the central, primary black hole. The naked black hole will then sink by dynamical
friction on a longer timescale \citep{callegari09,callegari11,khanminorlucio12}. 

\begin{figure}\sidecaption
\hbox{
  \includegraphics[width=0.65\textwidth]{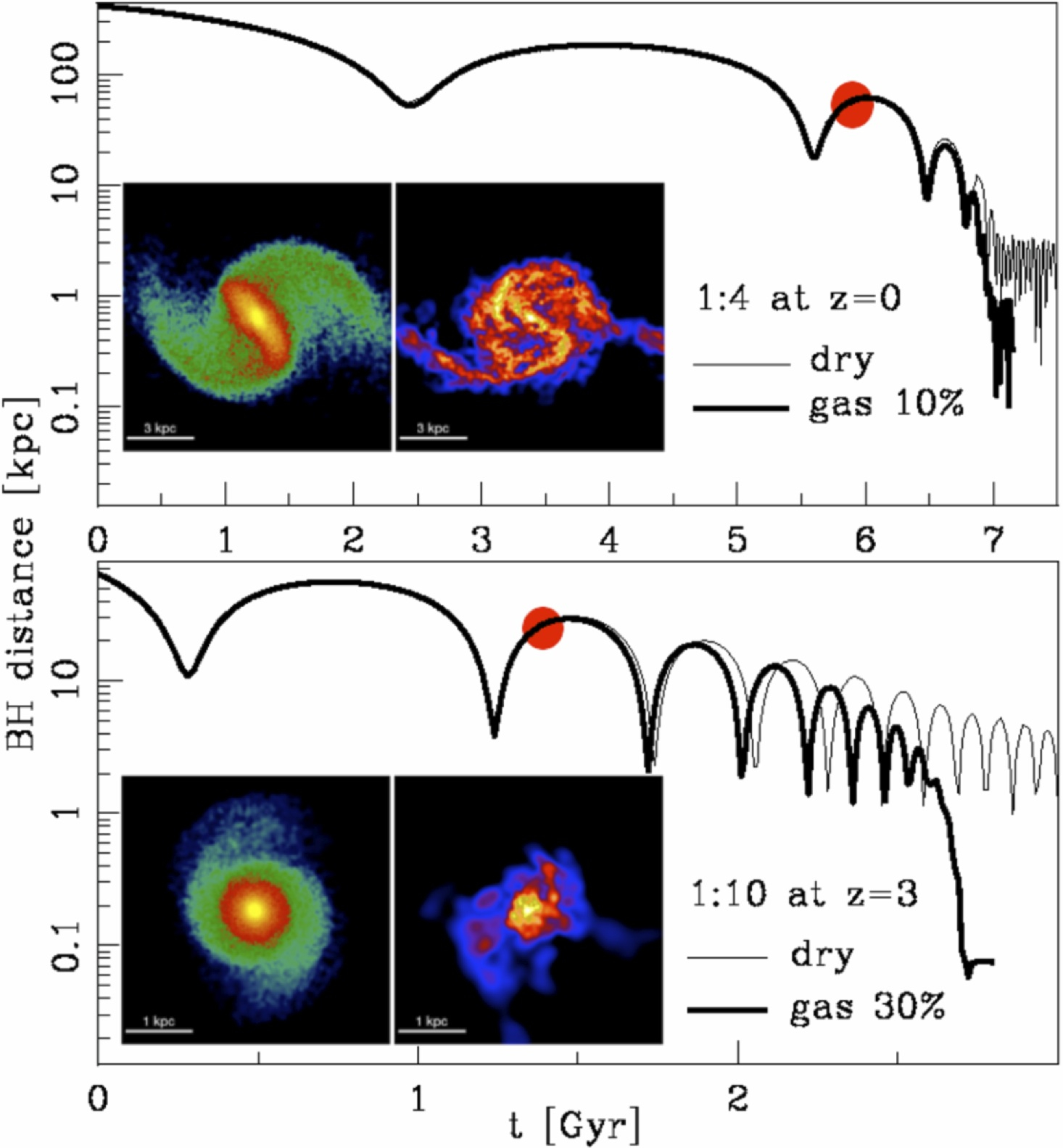}}
\caption{Upper panel: black hole separation as a function of time for a 1:4 merger.
  The thin and thick lines refer to the dry (gas free) and wet (with gas fraction of 10\%) cases, respectively. The
  inset shows the color-coded density of stars (left) and gas
  (right) for the wet case at $t=5.75$ Gyr (marked with a red dot on the
  curve); each image is 12~kpc on a side, and colors code the range
  $10^{-2}-1\,\msun $~pc$^{-3}$ for stars, and
  $10^{-3}-10^{-1}\msun$~pc$^{-3}$ for the gas.
  Lower panel: black hole separation as a function of time for a 1:10 merger (upper panel). The thin and
  thick line refer to the dry and wet (with gas fraction of 30\%)
  cases, respectively.  
  The inset shows density maps at $t=1.35$~Gyr for the wet merger:
    images are 4~kpc on a side (color coding as in upper panel). Courtesy  of \cite{callegari09}.
}
\label{fig:callegari1}
\end{figure}

Higher-resolution simulations of disc galaxies have recently revealed the occurrence of additional features, indicating  how rich is the outcome of
mergers under different initial conditions. 
Van Wassenhove et al. (2014) have shown that, 
as the gas-rich merger progresses, the newly formed stellar nucleus of the less massive galaxy, denser on small scales, is able to dissolve the
less concentrated nucleus of the primary, via impulsive tidal heating. This is illustrated in Figure~\ref{fig:vanvasse} which shows  how 
the denser nucleus of the secondary, at the end of the merger, 
finds itself in the midst of the mass distribution, having dissolved the nucleus of the main galaxy. 

\begin{figure}
\hbox{
  \includegraphics[width=0.33\textwidth]{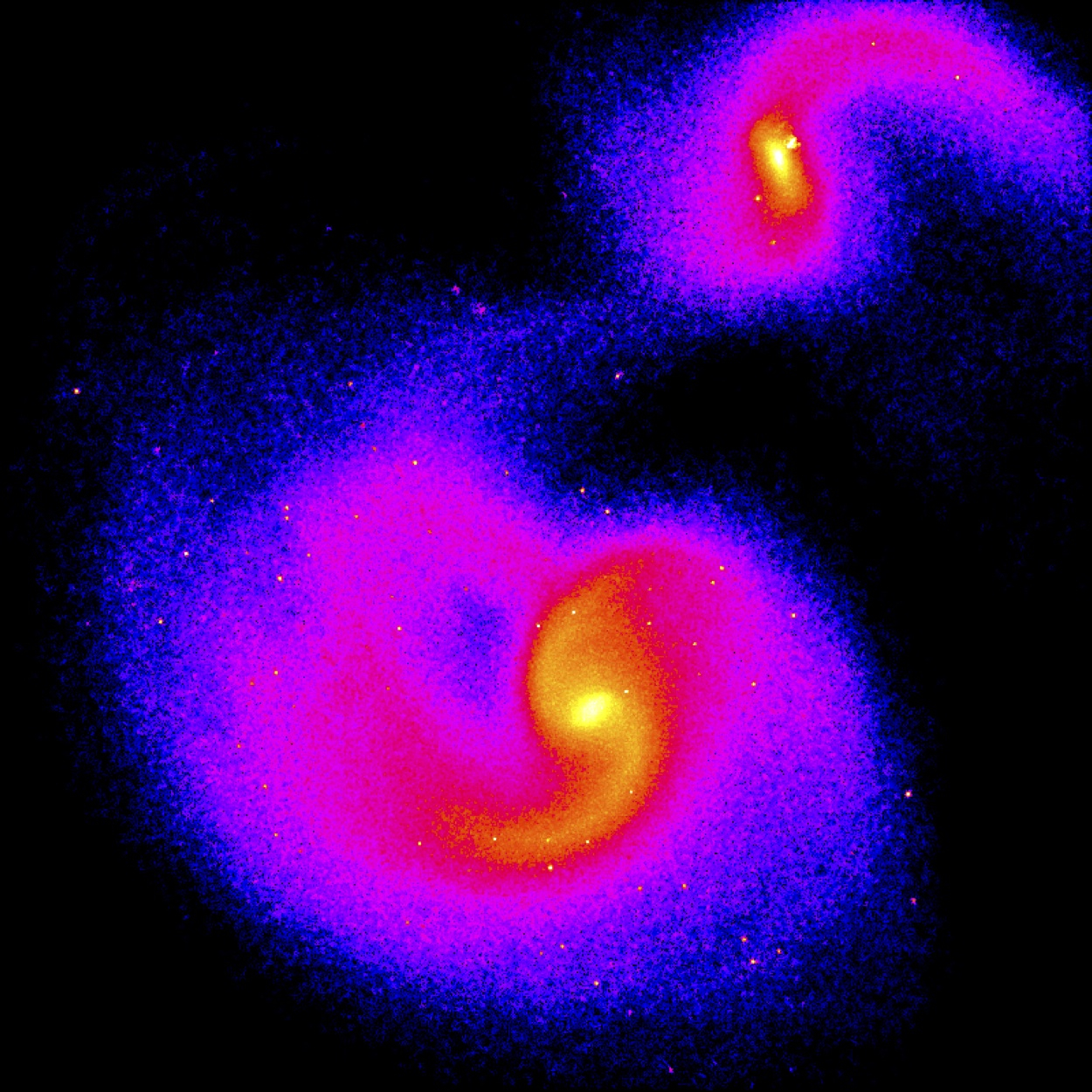}
  \includegraphics[width=0.33\textwidth]{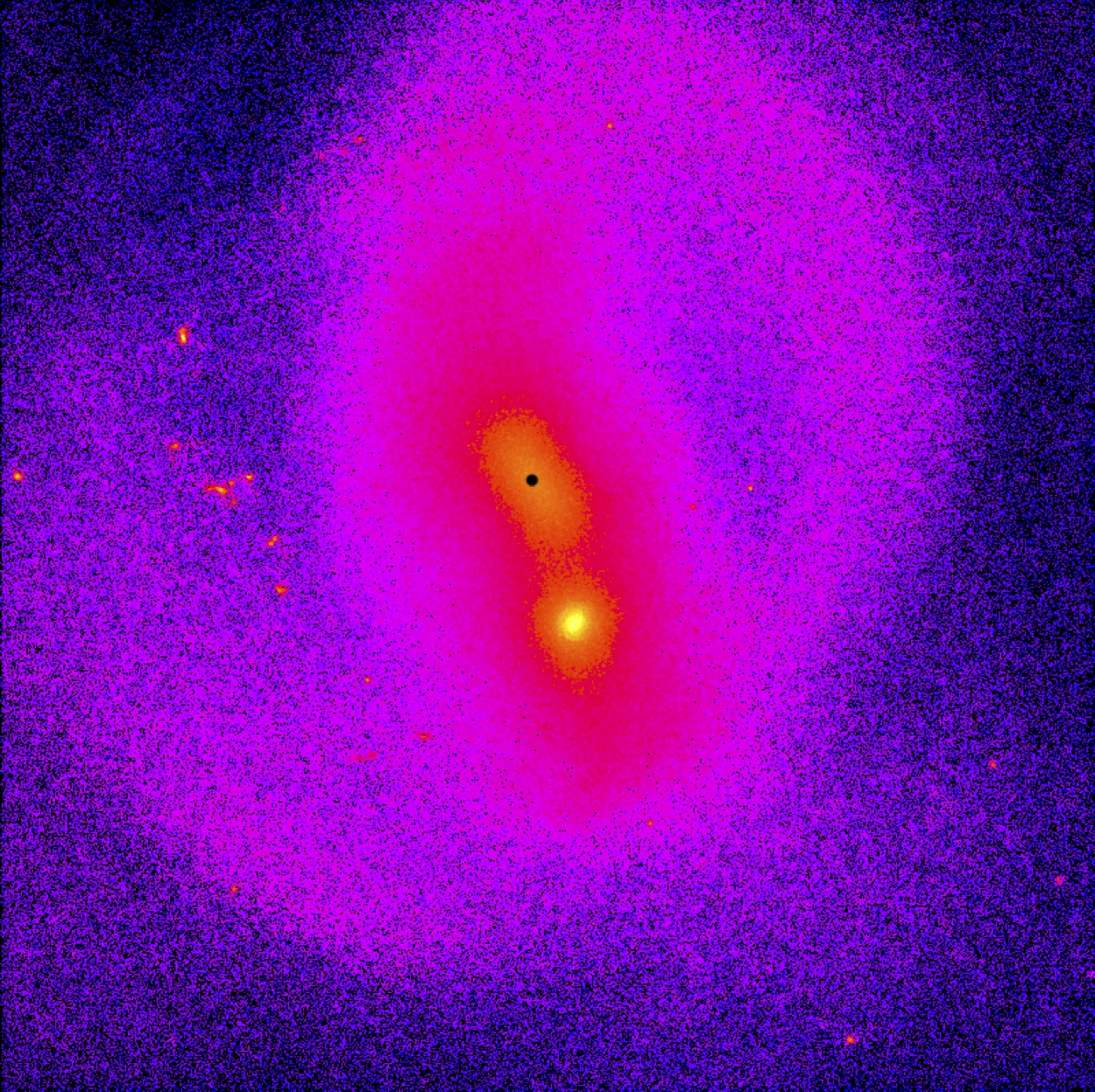}
   \includegraphics[width=0.33\textwidth]{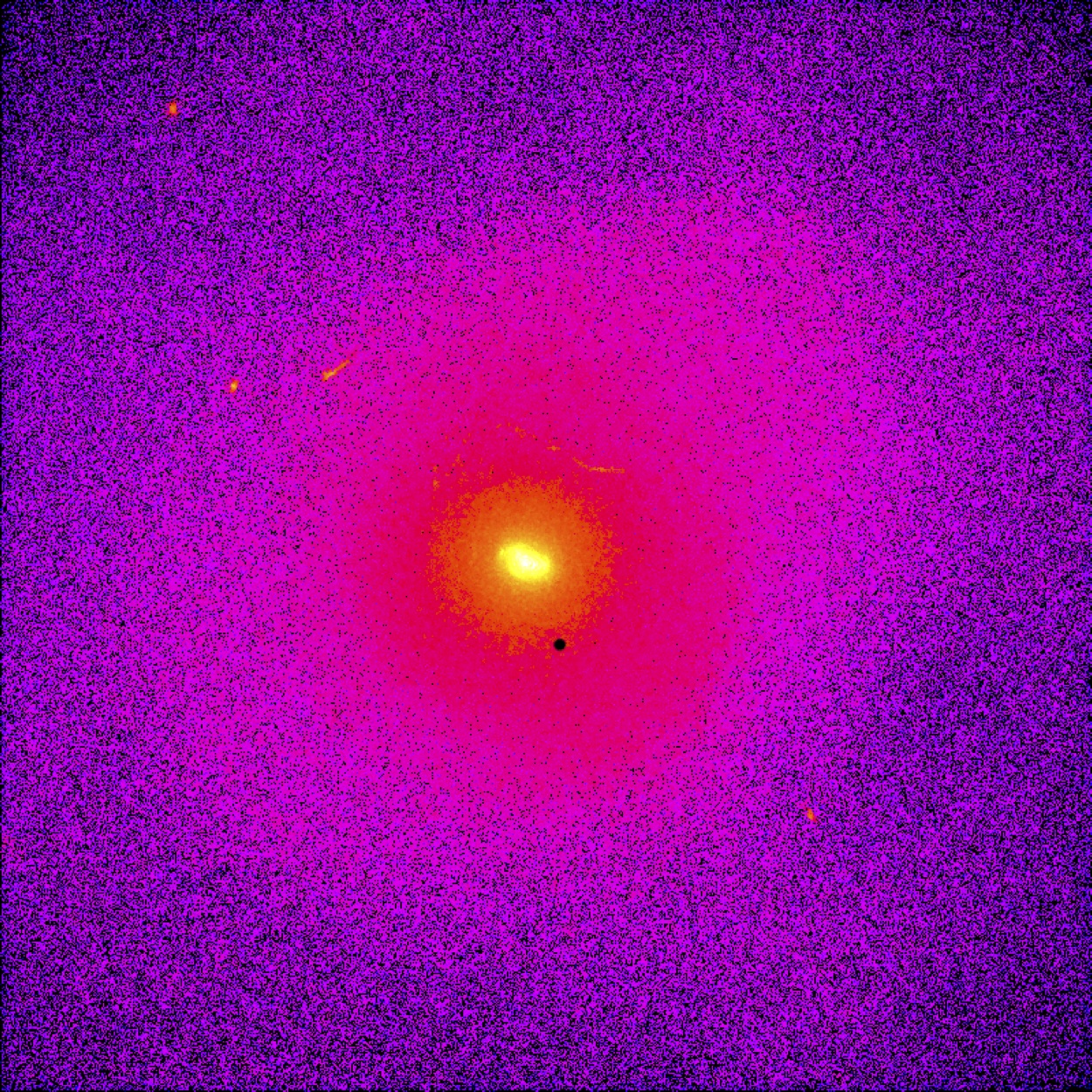}}
\caption{
Time sequence of  stellar density snapshots in the 1:4 coplanar, prograde-prograde merger after the second peri-centre passage, at times 1.2, 1.43, and 1.48 Gyr, respectively.  The scale of the left and central snapshots is 8 kpc, and 2 kpc for the right snapshot. A black dot marks the black hole in the primary galaxy nucleus which is dissolved during the interaction, while the secondary is at the centre of the highest density region of the secondary galaxy. In the last panel the secondary nucleus and black hole are near the
centre of the mass distribution.  Courtesy of  \cite{vanwasse14}.
}
\label{fig:vanvasse}
\end{figure}

\subsection{Black hole pairing in minor mergers: the role of mass accretion}
\label{sec:minor}
Minor mergers among galaxies with mass ratios 1:10 or less show behaviours that are extremes, along the 
sequence of unequal-mass mergers, and may lead to wandering black holes, even in presence of
a sizeable fraction of cold gas \citep{callegari09,callegari11}. The fate of black holes in minor mergers depends not only on the gas content but on the orbital parameters,
such as the degree of co-planarity, 
and in addition, on new input physics (neglected for seek of simplicity), i.e. accretion.
During the encounter
the secondary black hole as well as the primary can accrete from the surrounding gas and increase their mass.
A mass increase 
can influence the dynamics of the secondary black hole, as a larger mass implies a more rapid sinking by dynamical friction. This correlation has been found in
a number of simulations by Callegari et al. (2011)  who showed that the black hole mass ratio $q$ is not conserved during
the merger.  The secondary black hole is subjected to episodes of accretion 
which enhance the mass by an order of magnitude when interacting with the gas of the primary galaxy.
Thus, the black hole mass ratio does not mirror that of the galaxies, and can be 
much higher than the initial value indicating that black holes in unequal-mass mergers may carry comparable masses at the time they form 
a close pair.

Figure~\ref{fig:callegari2} summarises these findings, i.e.  the correlation between the ability of pairing (measured evaluating the black hole
relative separation) 
and the mass ratio $q,$ evaluated at the end of the
simulation.  Coplanar prograde mergers with higher fractions of gas lead to higher $q$ and smaller black hole separations.  Inclined
mergers with large gas fractions can instead fail in bringing the black holes to a small separation.
Torques acting on the satellite during the early phases of the merger are weaker for higher inclinations, and for this reason the increase in mass ratio $q$ during the first three orbits is milder than in the coplanar case with the same gas fraction.
Moreover, a higher inclination corresponds to a slower orbital decay so that the satellite galaxy undergoes a larger number of tidal shocks 
before being disrupted, preventing further episodes of substantial accretion onto the secondary black hole. 
Finally, gas-rich mergers on closer orbits (i.e. with smaller peri-centre) are less effective in pairing contrary to what expected.
Because of the smaller distance of approach and higher relative velocities between the satellite and the surroundings, ram pressure strips gas
effectively, reducing the importance of the starburst that made the satellite less susceptible to stripping, and the accretion process onto
the black hole.  The joint action of  these effects is therefore conducive to weak pairing, irrespective of the large amount of gas
present initially. 

\begin{figure}\sidecaption
\hbox{
  \includegraphics[width=0.65\textwidth]{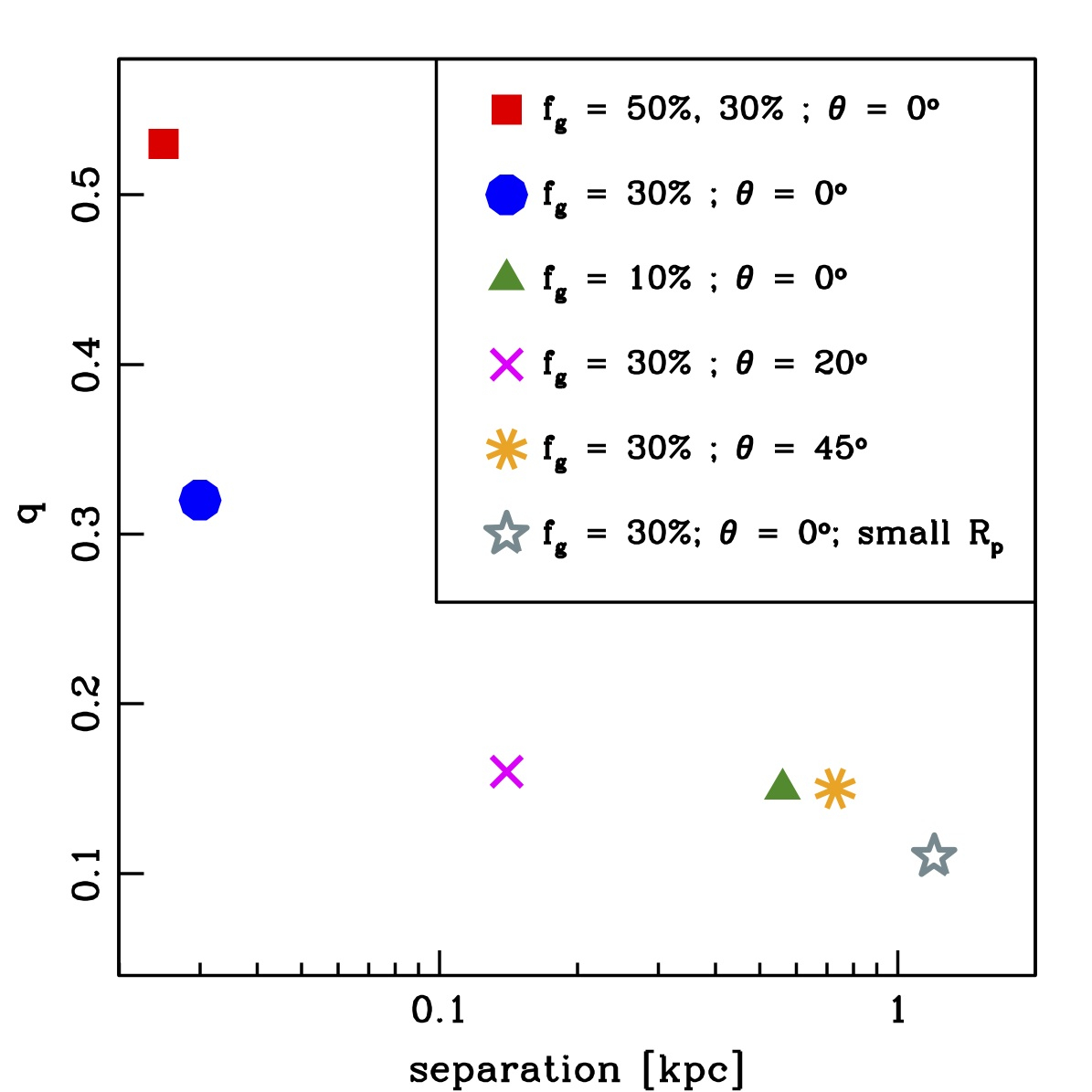}}
\caption{Black hole mass ratio $q$ versus relative separation, at the end of the simulation (when either a pair forms 
on the scale of the force resolution (10 pc), or the secondary black hole wanders at the periphery of the main galaxy), for the 1:10 mergers explored in \citep{callegari11}, labelled
  according to their initial gas fractions $f_{\rm g}$, orbital
  inclination $\theta$ and initial peri-centre $R_{\rm
    p}$. Courtesy of \citep{callegari11}.}
\label{fig:callegari2}
\end{figure}

In summary, minor mergers appear to fail in forming close black hole pairs in a number of cases, as the less massive galaxy
is disrupted by tidal and ram pressure stripping at earlier times during the encounter so that dynamical friction is unable to deliver the secondary black hole to the centre of the main galaxy, within a Hubble time. 
The boundary between failure and success, i.e. between coalescence and wandering, is still poorly determined as it depends on the geometry, gas content and internal structure  of the galaxies, and on the follow-up black hold dynamics on smaller scales \citep{khanminorlucio12}. 
Direct $N$-body simulations of gas-free minor mergers 
have shown that black hole coalescences can occur on timescales of one to a few Gyrs, regardless the mass ratio provided that  its value
$q\simgreat q_{\rm crit}\sim  0.05-0.1$ \citep{khanminor12}. The rather abrupt transition at $q_{\rm crit}$ appears to
result from the monotonic decrease of merger-induced triaxiality in the main galaxy with decreasing mass ratio.
The secondary galaxy is too small and light to significantly perturb the massive primary, slowing down the rate of binary
single star interactions and hardening. 
Judging from the results of simulations of galaxy minor mergers from a limited sample comprising gas-poor and gas-rich cases, 
a very rough boundary between coalescence and 
wandering appears to be at  $q_{\rm crit}\simless 0.1$.
Along parallel lines,  $N$-Body/SPH  cosmological simulations of massive disc galaxies, inclusive of black hole seed formation and growth, have shown that satellite galaxies containing black hole seeds are often tidally stripped as they merge with the primary 
while building the main galaxy disc. This creates naturally a population of wandering middleweight black holes in the massive spiral (from 5-10 wanderers), remnants of satellite cores \citep{bellovary10}. 

\section{Black hole dynamics in gaseous nuclear discs}
\label{sec:nucleardisc}

SPH simulations of black hole dynamics in massive, rotationally supported nuclear
discs represent a benchmark for 
studying the process of binary formation and coalescence, in gas-rich environments \citep{escala05,dotti06,dotti07,dotti09,fiacconi13}. 
These are {\it not ab initio} simulations, since 
the disc, in rotational equilibrium, is already in place as part of
the remnant galaxy, or of the main galaxy, in case a minor merger has delivered the secondary black hole inside the disc
of the massive host. 

At present, there is no analytical model nor simulation that can trace the black hole dynamics in nuclear 
discs from the disc's periphery (at $\simless 100$ pc) down
to the $\simless 10^{-3}$ pc scale (i.e. close and below $a_{\rm gw}$) where gravitational waves drive the inspiral, due 
to the susceptibility of the gas to undergo gravitational instabilities 
conducive to star formation episodes and to the complexity
of the gas thermodynamics in neutral and ionised media present on the smallest scales.
In this context, key elements are the {\it rotation} of the underlying background, its {\it self-gravity}, 
the degree of {\it gas dissipation}, and the nature of {\it turbulence and viscosity}. 
Large scale gas discs  can cool down, develop turbulence and inhomogeneities in the form of 
massive clumps which become sites of star formation.  Gas can also dissipate the 
kinetic energy of the moving black holes via radiative cooling in a disc on smaller scales when the disc around the binary is nearly Keplerian. 
Thus,  black hole inspiral in gaseous discs is mainly governed 
by processes of angular momentum exchange and radiative cooling. A compelling question to pose is whether angular momentum transport  resulting from
the gravitational interaction of the black holes with the gas is faster than that from the slingshot of stars. 

In gaseous discs, we are led  to distinguish three phases.
There exists an early phase I-g of {\it nuclear-disc-driven migration} 
during which non axisymmetric perturbations in the density field excited by the 
gravitational field of the black hole(s) cause the braking of the orbit in regions where the disc dominates the gravitational potential {\citep{escala05,dotti06,dotti07,dotti09}.  The typical scale covered by I-g is between 100 pc down to $\sim 0.1$ pc.

With time, the gas mass enclosed in the orbit decreases below $\mt$ and the black hole dynamics is
dominated by their own gravitational potential. The binary then forms a Keplerian system.
This corresponds to the onset of phase II-g  of {\it binary-disc-driven migration.} 
In phase II-g, the tidal torques exerted by the binary on the disc
are sufficiently intense to repel gas away from the binary clearing a cavity, called {\it gap}
 \citep{farris14,rafikov13,hayasaki13,orazio13,roedig12,kocsis12,shikrolik12,noble12,roedig11,cuadra09,hayasaki09triple,
macfadyen08,hayasaki08,hayasaki07,ivanov99,gould00}. 
The binary is then surrounded by a {\it circum-binary} disc. Rotation in the circum-binary disc is nearly Keplerian but the disc's structure 
is affected by the binary, acting as a source of
angular momentum.  In phase II-g, black holes migrate under the combined action of viscous and gravitational torques which 
ultimately drive the binary into the third phase III of 
{\it gravitational-driven inspiral} where loss of orbital energy and angular momentum is due to the emission of
gravitational waves. Not for all black hole masses this gas-assisted inspiral leads to
coalescence in a Hubble time and more work is necessary along these lines  \citep{cuadra09}. 

Below, we  explore phase I-g 
considering  first a smooth nuclear disc and later a clumpy nuclear disc to study black hole migration on pc-scales. In a second step we will explore phase II-g when a circum-binary disc forms which controls the evolution of the binary on smaller scales.

\subsection{Nuclear-disc-driven migration}
\label{sec:migration}

\subsubsection{Smooth circum-nuclear discs}
\label{sec:smoothmigration}

In a number of targeted studies, the massive nuclear disc is 
described by a Mestel model: the disc, self-gravitating and axisymmetric, has rotation velocity $V_{\rm rot}$ independent of radius $R$.
With constant $V_{\rm rot}$, fluid elements in the disc are in differential rotation with $\Omega=V_{\rm rot}/R$, and
are distributed following a surface density profile $\Sigma(R)=\sigmadisc [\rdisc/R]$, where $\rdisc$ is a scale radius.
The disc mass within a radius
$R$ is then given by $M_{\rm Mestel}(R)=\mdisc [R/\rdisc]$, with $\mdisc=2\pi \rdisc^2\sigmadisc$, and the circular velocity
$V^2_{\rm rot}=G\mdisc/\rdisc.$  The disc
is pressure supported vertically, with aspect ratio $h/\rdisc$ of $\sim 0.1-0.05$, and isothermal sound speed $c_{\rm s}$ such that the Toomre parameter $Q$ is $\simgreat 3$ everywhere, to prevent the development of
gravitational instabilities.
The disc is embedded in a more massive Plummer stellar sphere, representing the innermost region of the galactic bulge:
hereon we refer to this configuration as circum-nuclear disc \citep{escala05,dotti06,dotti07,dotti09}. 

In this smooth  background (guaranteed by the large $Q$) one can trace the black hole dynamics, assuming a primary 
black hole of mass $\mbhuno$ at rest in the centre of the 
circum-nuclear disc, and a secondary black hole of mass $\mbhdue$ initially moving on a wide eccentric co-planar orbit
inside the disc.  The mass ratio $q$, the initial 
binary orbital elements, and 
  the disc mass $M_{\rm Mestel}$ enclosed in
the binary orbit (in excess of the binary mass $\mt$, in the simulated volume), 
are free parameters 
 to mimic different encounter geometries and mergers of galaxies with
different stellar/gas mass contents.  

Assisted by a series of $N$-Body/SPH simulations, these studies have highlighted key differences in the black hole dynamics
compared to that in spherical collisionless backgrounds, the most remarkable being 
the dragging of the moving black hole into a co-planar co-rotating orbit with null eccentricity before the black holes form a binary
\citep{dotti06,dotti07,dotti09}.

The simulations show that any orbit with large initial eccentricity is forced into circular rotation in the disc. In the different panels of Figure~\ref{fig:smoothdottilike}, we show the over-density excited by the black hole  
along the orbital phase, for an initial value of the eccentricity $e_0$ equal to $0.9$.
The wake, that in a uniform medium, 
trails the motion of the perturber maintaing its shape, here changes both orientation and shape,
due to the differential rotation of the underlying disc. 
The wake is trailing behind when the black hole is at pericentre, but is leading ahead when at apocentre,
since there, the black hole is moving more slowly than the background, and this causes
a temporary acceleration on $\mbhdue$. It is important to remark that 
this is a rapid change of $e$ occurring on few orbital times.  Circularisation is a fast process, and it is faster the cooler is the
disc, i.e. the denser is the disc. Sinking times are found to depend on the equation of state adopted, i.e
on the polytropic index $\gamma$ used to model the thermodynamic behaviour of the gas.
\begin{figure}
\includegraphics[width=1.0\textwidth]{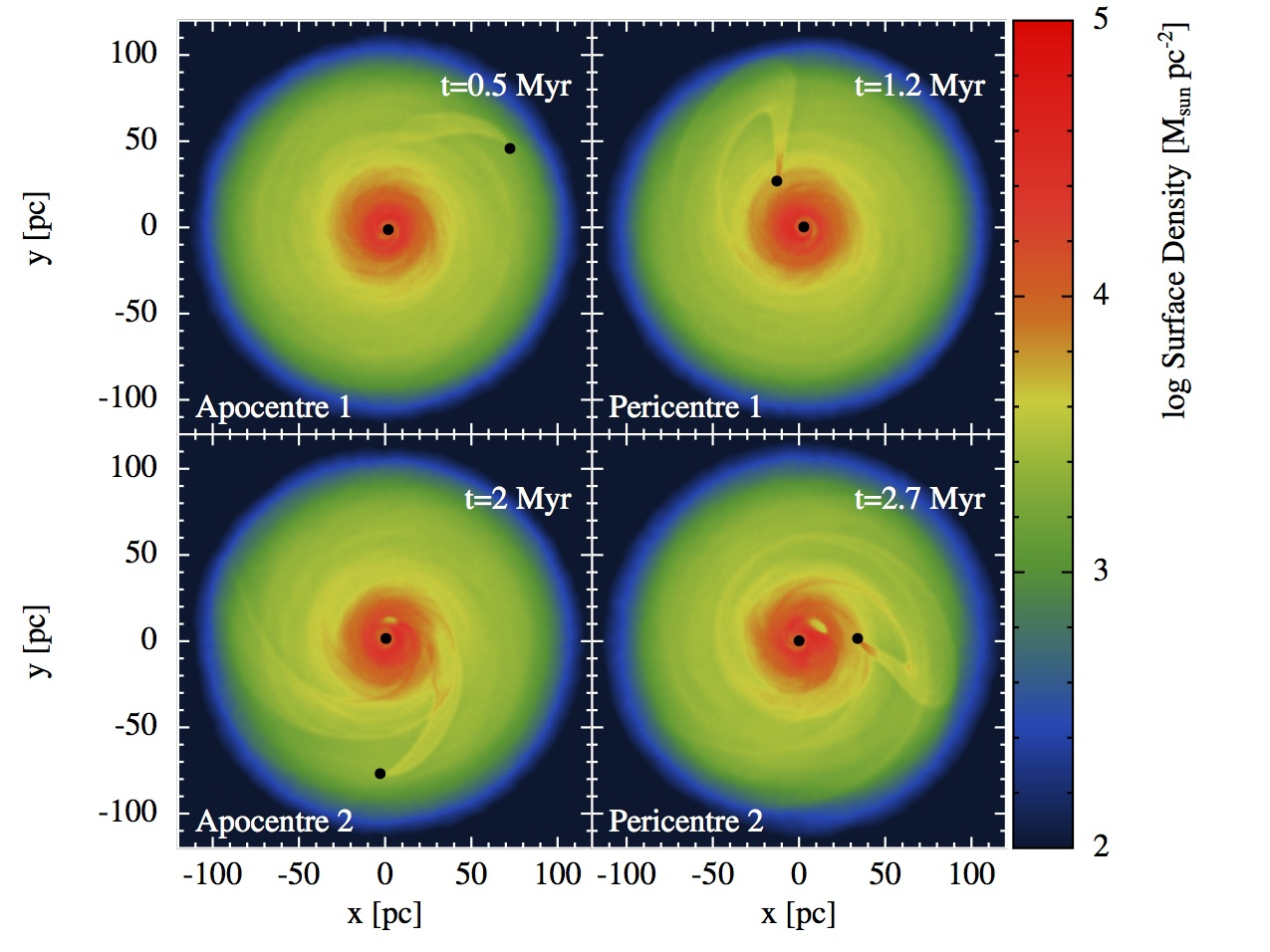}
\caption{
Colour-coded gas surface density of a smooth disc of mass $\mdisc=5\times 10^8\,\msun$,  for a black hole binary with 
$q=0.1$ and a primary black hole of $10^7\,\msun$. The initial eccentricity  is $e_0=0.7$.  Snapshots refer to four different times, covering the process of circularisation lasting a few Myrs.  The gas and the secondary black hole rotate counter-clockwise.
The position of the black holes is marked by black dots.  In the top and bottom left (right) panels the density wake excited by the secondary black hole 
is leading (trailing) the orbit, resulting in the circularisation of the relative orbit.
Courtesy of D. Fiacconi. }
\label{fig:smoothdottilike}
\end{figure}

A further signature of a rotating background is
the {\it angular momentum flip} of an initially counter-rotating
orbit, if it exists \citep{dotti09}.\footnote {This is a possibility that may occur in the case of a minor merger
where the incoming black hole in the satellite galaxy enters the main galaxy from a
co-planar counter-rotating orbit.}  Initially, the gas opposes to the motion of the black hole as the density perturbation is always
in the form of a trailing wake which causes an effective brake.  The change of the orbital angular
momentum from negative to nearly null values  is further facilitated by the fact that, while the orbit
decays, the black hole interacts with progressively denser regions of
the disc. The orbit is nearly radial when the orbital angular
momentum changes sign but the change is so rapid, relative to the orbital time, 
that the black hole is forced to co-rotate. When co-rotation establishes along an eccentric orbit, the orbital
momentum increases under the circularising action of dynamical
friction (non axisymmetric wake)  in its co-rotating mode.   Thus, a further
prediction of black hole inspiral in rotating discs is that
gas-dynamical friction conspires to turn counter-rotating orbits into 
co-rotatating  ones, even before the formation of a  Keplerian binary.

After circularisation, 
the secondary black hole continues to spiral in as it experiences a net negative torque, despite having reduced 
its relative velocity with respect to neighbouring fluid elements.\footnote{In a uniform, isotropic gaseous background,
gas-dynamical friction vanishes when the velocity of the perturber falls below the sound speed \citep{ostriker99}. \cite{kim08}
extend the work by \cite{ostriker99} considering 
double perturbers moving on a circular orbit in a homogeneous gaseous medium. They find 
that the circular orbit makes the wake of each perturber asymmetric, creating an over-dense tail at the trailing side. The tail not only drags the perturber backward but it also exerts a positive torque on the companion. This finding shows that the orbital decay of a perturber in a double system, especially in the subsonic regime, can take considerably longer than in isolation.} 
Dynamical friction
is a non-local process and in a disc there is a residual velocity difference
between the black hole and the more distant rotating fluid elements. One can view the migration process described in the text again as a manifestation
of the larger scale gravitational perturbation excited by the black hole, but this time the drag
is inside a rotating inhomogeneous background. The net torque results
from the sum of positive (inside the black hole orbit)  and negative (outside) contributions as the perturbation is highly non axisymmetric due to differential rotation.} 
The black hole $\mbhdue$ is able to 
excite a non axisymmetric perturbation in the disc structure which produces a net negative torque on $\mbhdue$.  This process is reminiscent to Type I planet migration, but with key differences.
While in planet migration, the central star dominates the gravitational potential and the disc's self-gravity is negligible, 
in disc-driven black hole migration (phase I-g)  the disc is dominant, while the gravity of the central black hole is negligible.  In Type I migration, the net torque on the planet is the sum of the Lindblad and co-rotating resonances, computed in
the linear perturbation theory under the hypothesis that the planet migrates on a timescale much longer than the orbital time,
so that the small-amplitude perturbation is periodic in the disc frame.
By contrast, during phase I-g of black hole migration, the torque on the secondary black hole comes from the 
non-linear density perturbations that $\mbhdue$ excites in the disc. Figure~\ref{fig:J-migration} shows the fast orbital decay that the secondary black hole experiences after circularisation \citep{fiacconi13}.

\begin{figure}\sidecaption
\includegraphics[width=0.70\textwidth]{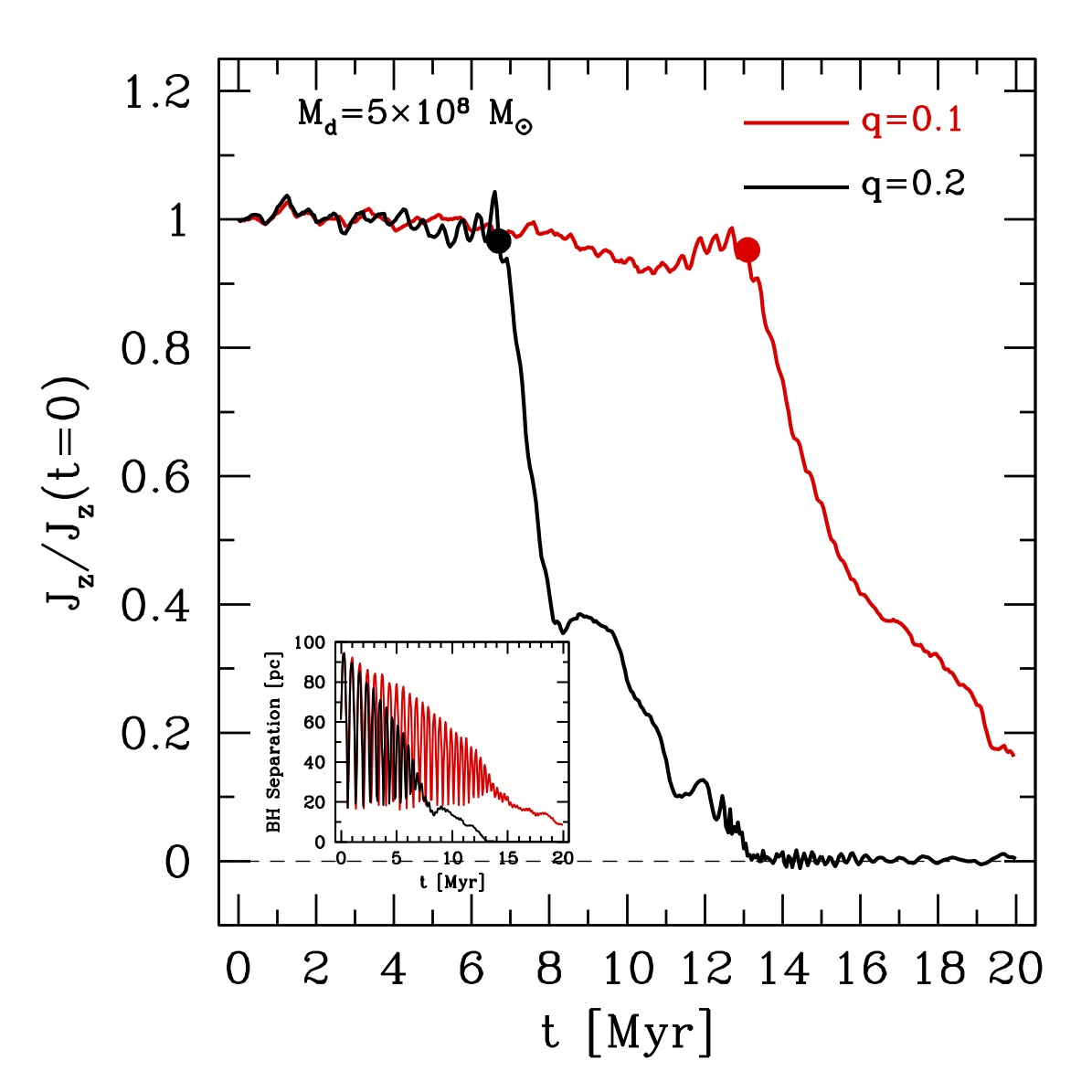}
\caption{The evolution of the angular momentum of the secondary black hole 
orbiting inside a smooth circum-nuclear disc of mass $\mdisc=5\times 10^8\,\msun$. 
The initial eccentricity is $e_0=0.9.$ Red and black colours refer to $q = 0.1$ and $q = 0.2$, respectively ($\mbhuno=10^7\,\,sun$ in the runs). The dot  marks the time at which the  circularisation of the orbit by
dynamical friction is completed. The inset shows the black hole separation (in pc)  versus time (in Myr). At circularisation, 
$M_{\rm Mestel}(a)/\mbhdue\sim 150$ and 75 for the two cases, respectively.   Courtesy of D. Fiacconi.}
\label{fig:J-migration}
\end{figure}

To gain some
insight into nuclear-driven-disc migration in a Mestel disc, we estimate the migration time  following a simple argument 
to capture key dependences of $\tau_{\rm mig}$ on the disc properties  and black hole mass \citep{armitagebook13}. 
In the two-body approximation, a fluid particle, approaching $\mbhdue$ along a straight-line with impact parameter $b$
and relative velocity $ { \rm v}_{\rm rel}\sim a\delta \Omega\sim b\Omega$ experiences a velocity change parallel to ${\bf v}_{\rm rel}$
of the order of $\delta v_{||}\sim 2G^2\mbhdue^2/(b^2{\rm v}^3_{\rm rel}),$ 
where $a$  denotes the black hole distance from the centre of the disc. As gas exterior to the secondary black hole
moves more slowly than the black hole in the disc, it gains velocity parallel to ${\bf {\rm v}}_{\rm rel}$ increasing its angular momentum. 
As angular momentum is conserved,  this implies a  {\it decrease} in the angular momentum  per unit gas mass for the black hole 
equal to $\sim - a \delta {v}_{||} $.  Note that gas interior to the black hole orbit  exerts a torque of opposite sign,
so that the net torque depends on a delicate balance.  As simulations show inward migration and 
larger torques in the black hole vicinity, we compute the rate of change of angular momentum
considering only neighbouring gas particles in the trailing side of the spiral density perturbation, contained in a cylinder of scale height  $b\sim h$.  Accordingly, the mass flux on $\mbhdue$ is  $\delta m/\delta t\sim \sim 2\pi h \Sigma  {\rm v}_{\rm rel}\sim 2\pi h^2\Sigma \Omega$, 
where $\Sigma$ and $\Omega$ are evaluated at $a$. 
The resulting torque on the black hole can thus be written as $T_{\rm I}^{\rm Mestel}\sim -4\pi \zeta [\mbhdue/M_{\rm Mestel}(a)]^2
\Sigma a^4\Omega^2,$ where  $\zeta=\zeta'(a/h)^3$ brackets uncertainties in the normalisation of the torque and its
dependence on the aspect ratio. 
As in a Mestel disc the circular velocity is independent of radius $a$, the black hole sinks from an initial radius $a_{\rm i}$ to a much smaller radius $a_{\rm f}$ on a 
migration time scale 
$\tau^{\rm I}_{\rm mig, Mestel}\sim C \Omega_{\rm i}^{-1} [M_{\rm Mestel}(a_{
\rm i})/\mbhdue]$, where $C=(h/a)^3/(4\zeta').$
 \footnote{The coefficient $\zeta$  
can be inferred from dedicated numerical experiments. In the case explored,  
the coefficient $\zeta'\sim 0.04$, to match the sinking time with a simulation. A systematic analysis is necessary to
estimate $\zeta'$ in a Mestel disc (paper in preparation). Furthermore, the scaling of $\tau^{\rm I}_{\rm mig, Mestel}$ with the aspect ratio $h/a$ can not be derived
from this elementary argument, as discussed in \cite{armitagebook13}.} 
We remark that the scaling of $\tau^{\rm I}_{\rm mig, Mestel}$ with the disc and black hole mass holds true provided $M_{\rm Mestel}(a)>\mbhuno>\mbhdue.$

Nuclear discs, stable against fragmentation, are nevertheless ideal. The gas can not be treated as a simple one-phase fluid.
Galactic discs are sites of local gravitational instabilities conducive to star formation episodes:  massive stars inject energy
in the form of winds and supernova blast waves, feeding back energy into the disc and the gas is multi-phase, and clumpy.
Thus, it is of importance to understand how the black hole sinking is affected by the inhomogeneous substructure of star forming discs.
To this aim, in the next subsection, we explore black hole dynamics in clumpy discs.

\subsubsection{Clumpy circum-nuclear discs and stochastic orbital decay}
\label{sec:clumpydisc}

In
real astrophysical discs, massive gas clouds coexist with
warmer phases and a polytropic equation of state, often used in SPH simulations, provides
only an averaged representation of the real thermodynamical
state.  Cold self-gravitating discs are unstable to fragmentation and
attain stability when stars, resulting from the collapse
and/or collision of clouds, inject energy in the form of winds and
supernova blast waves, feeding back energy into the disc now composed
of stars and a multiphase gas.  Due to the complexity of implementing
 this rich physics at the required level of accuracy,
a first step ahead is to insert a phenomenological cooling prescription to allow the formation of clumps
in a controlled way \citep{fiacconi13}. 
Clumps of size $\sim 5$ pc in the mass interval between $10^5\,\msun$ and $10^7\,\msun$ develop in the disc, and evolve as they mass segregate, collide with each other and interact with the secondary black hole.
Acting as massive perturbers, they disturb the otherwise smooth black hole orbital decay due to the stochastic behaviour 
of their torques that are not coherent in time.
Several close encounters between $\mbhdue$ and the massive clumps act as gravitational slingshots, causing an 
impulsive exchange of orbital energy and angular momentum. Thus, the black hole  deviates from its original 
trajectory either outwards, or inwards or out of the disc plane (above the typical
scale height of the disc). When moving on an inclined orbit  
the black hole 
experiences the weaker dynamical friction of the stellar background, resulting in a longer orbital decay timescale. 
The secondary black hole can also be captured by a massive clump forming a pair which segregates
rapidly toward the centre.  Figure~\ref{fig:stochastic-density} shows the evolution of the gas surface density of a selected run, and 
the position of the two black holes (marked as white dots) at four different times. 

\begin{figure}\sidecaption
  \includegraphics[width=1.0\textwidth]{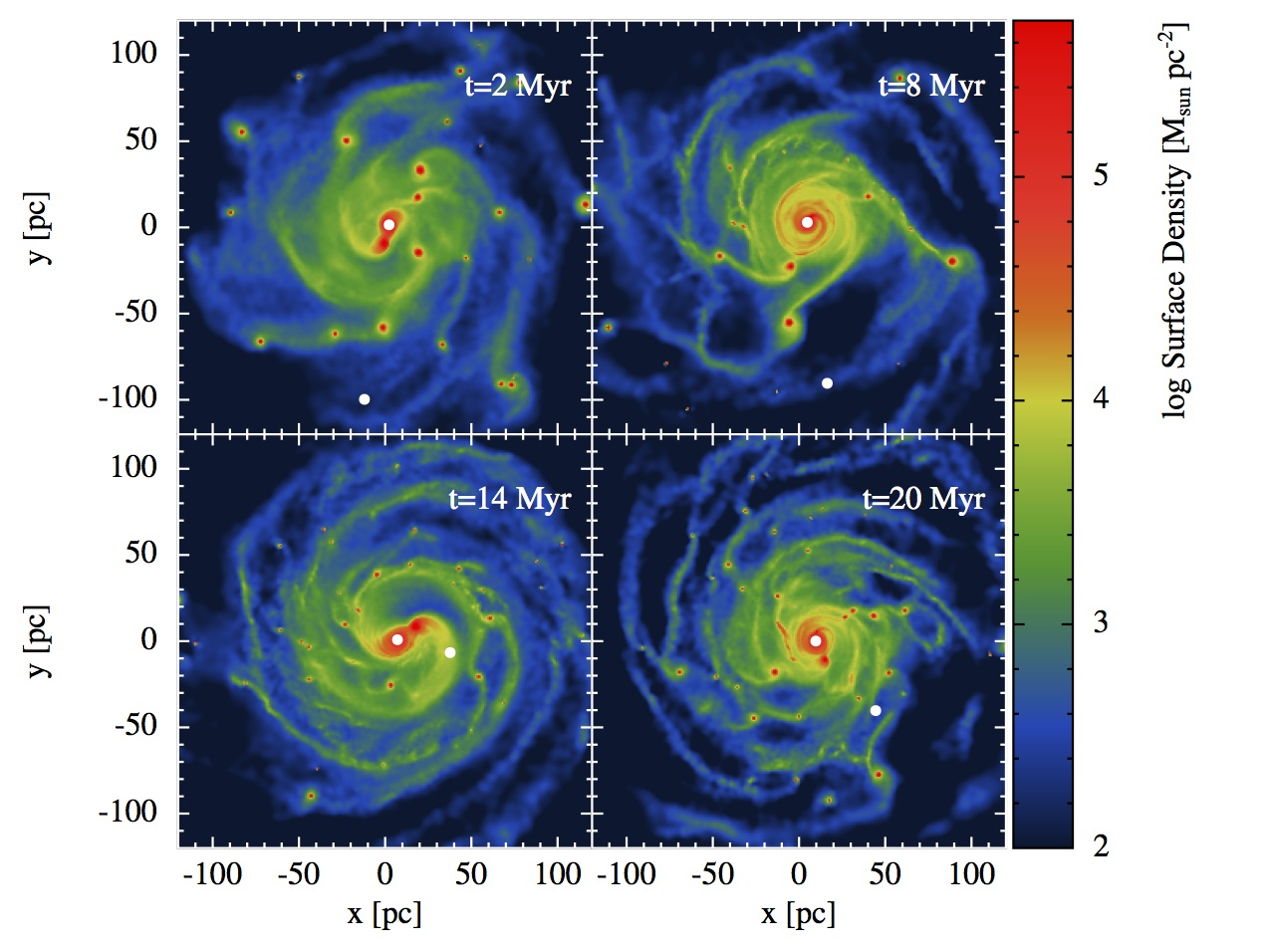}
\caption{Colour-coded face-on view of the gas surface density of a clumpy disc model with  $\mdisc=5\times 10^8\,\msun,$ 
$\mbhuno=10^7\,\msun$, $q=0.1$ and initial eccentricity $e=0.7$, plotted at four different times.  The position of the 
black holes is marked by white dots. Courtesy of D. Fiacconi.}
\label{fig:stochastic-density}
\end{figure}

The {\it stochastic} behaviour of the black hole orbit, resulting from the incoherence of torques, 
emerges mainly when the clump to
black hole mass ratio is $M_{\rm clump}/\mbhdue\simgreat 1$. This enlarges the values  of the decay time which now range from less than $\simless 1 $ up  to  $\simgreat 50$ Myr. This suggests that describing the cold clumpy phase of the interstellar medium in nuclear discs, albeit so far neglected, is important to predict the black hole dynamics. Ongoing simulations in a multi-phase star forming nuclear disc
resulting from the collision of two discs following a merger produce results that are intermediate between the smooth and clumpy 
case (Lupi et al. in preparation).

\subsection{Binary-disc-driven migration}
\label{sec:gap?}

In Section~\ref{sec:smoothmigration} we followed the black hole inspiral during phase I-g, in presence of a self-gravitating, rotationally supported disc much heavier than the binary, a condition leading to migration as described in \citep{escala05,dotti06,dotti07,dotti09}.
However, with binary decay, the disc mass $M_{\rm Mestel}$ enclosed in the black hole orbit $a$ decreases 
with time falling below $\mt$.  The black holes then form a Keplerian binary surrounded by a 
less massive disc, called {\it circum-binary} disc, dominated by the gravity of the binary and its quadrupolar field.

If we impose continuity in the physical processes, there might exist an intermediate phase whereby migration of 
the secondary black hole is controlled by {\it resonant}  torques. In close resemblance to Type I planet migration, and for
very small binary mass ratios $q\ll 1$, the resulting torque on $\mbhdue$ is 
$T_{\rm I}^{\rm mig,Kep}=-\zeta_K[\mbhdue/\mbhuno]^2\Sigma a^4 \Omega_{\rm K}^2$, 
where  $\Omega_{\rm K}$ is the Keplerian
rotational velocity in the gravitational field of $\mbhuno$ evaluated in $a$, $\Sigma$ the disc surface density in the immediate vicinity of $\mbhdue$, 
$\zeta_K=(1.36+0.54 \alpha)(a/h)^2$, 
 and $\alpha$ the slope of the surface density profile $\Sigma\propto a^{-\alpha}$ of the underlying Keplerian disc \citep{tanaka02,armitagebook13}.
Notice that because of the natural scaling present in the problem, the torque  $T_{\rm I}^{\rm Kep}$ differs from the expression of  $T_{\rm I}^{\rm Mestel}$ having $\mbhuno$  in place of $M_{\rm Mestel}(a)$
as reference mass for the gravitational potential. In this case,  the migration time reads as  $\tau^{\rm I}_{\rm mig,Kep}\propto [\mbhuno/\mbhdue]
 [\mbhuno/M_{\rm disc}(a)]\Omega_K^{-1}$, under the condition that the disc mass enclosed in the 
 black hole orbit $M_{\rm disc}(a)<\mbhuno.$ 

As described in Section~\ref{sec:pairing}, 
black hole binaries form preferentially in the aftermath of major mergers so that the binary mass ratio $q\simless 1$.  Furthermore,
accretion drives $q$ to larger and larger values, in the case of minor mergers under specific circumstances \citep{callegari11}. Thus, 
it is quite likely that migration under the action of torques excited by resonances is a missing step, in the dynamical
evolution of black hole binaries.  At this time the binary is expected to alter profoundly  the structure of the underlying disc.

Thus, a key question  poses: will the black holes sink down to the domain of gravitational wave inspiral transferring their angular momentum to the disc, or would the binary stall?  Is there a phase II-g of migration and under which conditions?  This phase is in fact
somewhat controversial, as the black hole fate depends on whether the disc is a one-time, 
short-lived excretion disc \citep{lodato09,pringle91}, or an extended long-lived disc \citep{rafikov13}. 
These are conditions that are not recoverable from realistic larger-scale simulations as it is difficult to model the transition from a disc
dominated by self-gravity and gravito-turbulence to a disc dominated by magneto-hydrodynamical
turbulence stirred by magneto-rotational instabilities in the conducting fluid \citep{shikrolik12}.

Unless the binary is surrounded by a geometrically thick disc or envelope and decays promptly \citep{delvalle12,delvalle14}, the
tidal force exerted by the binary on the circum-binary disc is expected to eventually clear a cavity.
The picture is that the binary transfers orbital angular momentum to the disc by exciting non-axisymmetric density perturbations in the disc body, causing the formation of  a low-density, hollow region, called {\it gap} 
\citep{farris14,rafikov13,hayasaki13,roedig12,kocsis12,shikrolik12,roedig11,cuadra09,hayasaki09triple,
haiman09,macfadyen08,hayasaki08,hayasaki07,ivanov99,gould00,pringle91}. 
Viscous torques in the disc oppose gas clearing by the tidal field of the binary and ensure 
strong binary-disc coupling. Under these conditions, the binary enters a regime of slow orbital decay
[referred to as Type II migration in the case of planets \citep{artygap94,artyflow96,gould00,armitage02,armitagebook13}] during which the inner edge of the circum-binary disc 
compresses in coordination with the hardening of the binary, so that the size $\delta(t)$ of the gap decays remaining close to twice the
binary semi-major axis, $\delta \sim 2 a(t).$  

Due to the tidal barrier offered by the binary, gas piles up at the inner rim of
the disc.  One can view the binary as acting as a dam, halting the gas inflow. Accordingly, the accretion rate in the circum-binary disc is not constant in radius and
no strict steady state is ever attained in the disc body. In 1D modelling of circum-binary discs,  the disc 
is seen to evolve into a state of constant angular momentum flux   \citep{rafikov13}, and binary decay leads to
a secular and self-similar evolution of the disc as first suggested in \citep{ivanov99}. This 
happens when the system loses memory of the natural scale $a$, set by the size of the cavity and binary orbit, soon after the angular momentum injected by the binary has been transmitted to the larger scale extended disc  \citep{rafikov13}.
Gap opening implies longer hardening time scales compared to nuclear-disc-driven migration, now controlled by
the viscous time at the inner disc edge.

The migration time can be estimated as 
$\tau^{\rm II}_{\rm mig}\sim \tau_\nu [(\mbhdue+M^{\rm edge}_{\rm d})/M^{\rm edge}_{\rm d}]$
where $M^{\rm edge}_{\rm d}\sim \Sigma (R) R^2$ is the  local mass near the inner edge of the disc, at $R\sim 2a$-$3a$
where the surface density has a peak, 
 and $\tau_{\nu}$ the disc viscous time there: $\tau_\nu\sim (2/3)R^2/\nu\sim 2\pi R^2 \Sigma/{\dot M}$
 where $\dot M\sim 3\pi \nu \Sigma$ is the mass accretion rate in an unperturbed reference disc \citep{shakura73}.
 When $M^{\rm edge}_{\rm d}> \mbhdue$, the secondary black hole behaves as a parcel in the
 viscous disc and migrates on the viscous timescale, whereas when  $M^{\rm edge}_{\rm d}< \mbhdue$
 migration slows down and occurs of a timescale longer than $\tau_\nu$. Condition 
   $M^{\rm edge}_{\rm d}< \mbhdue$ is often referred to as secondary-dominated Type II migration \citep{haiman09,syer95}. The opposite regime is referred to as disc-dominated Type II migration.
   
  The timescale  $\tau^{\rm II}_{\rm mig}$ can be recovered if one assumes a torque on the binary of the form
   $T_{\rm II}^{\rm mig}\sim -\xi\, j_o{\dot M}\sim \xi'\,  j_o M^{\rm edge}_{\rm d}\Omega_K$ where $j_o=(\mu/\mt)(G\mt a)^{1/2}$ is the binary angular momentum per unit mass,
   and $\xi$ or $\xi'$ determined by numerical simulations, e.g. \citep{macfadyen08,roedig12,shikrolik12}.
   The above expression for the torque relates the
rate of binary orbital decay to the local disc mass $M^{\rm edge}_{\rm d}$ near the inner edge of the disc. 
Therefore, depending on how $\mdisc^{\rm edge}$ varies over the relevant  timescales, i.e. whether the disc is continuously re-filled of gas
to keep $\mdisc^{\rm edge}$ nearly stationary, or the disc mass is consumed before the binary has evolved substantially, orbital decay accelerates or decelerates, returning the problem to the old, outstanding issue on whether black holes are continuously fed in galactic nuclei or not, and on which
timescale.

\begin{figure}
  \includegraphics[width=0.75\textwidth,angle=90]{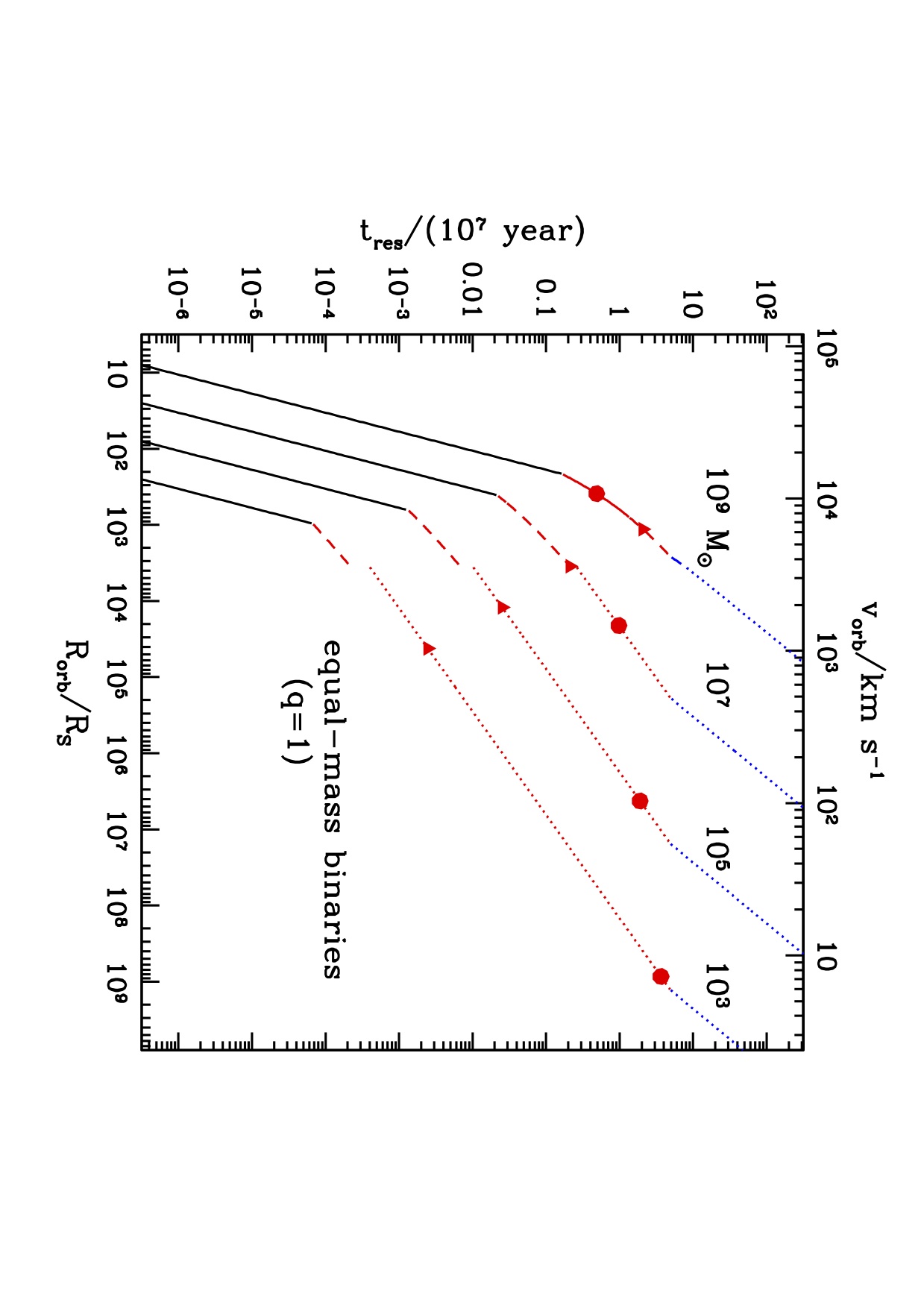}
\caption{Residence time $\vert a/{\dot a}\vert$ of equal-mass black hole binaries, embedded in a steady circum-binary disc, as a function of
the black hole separation (in units of $2G\mt/c^2$), as computed in \citep{haiman09}
for a reference disc model. The top $x$-axis label refers to the Keplerian relative orbital velocity of the black holes
in the binaries. 
The four curves correspond to binaries with total masses of $\mt=10^3,10^5, 10^7$ and $10^9\,\msun$  as labeled. The large dots denote the critical radius beyond which the assumed circum-binary Keplerian disc is unstable to fragmentation. Similarly, triangles denote radii beyond which the disc may be susceptible to ionisation instabilities (the gas temperature falls below $10^4$ K). In each case, blue/red colors indicate whether the disc mass enclosed within the binaryÕs orbit is larger/smaller than the black hole mass $\mbhdue$. The dotted/dashed/solid portion of each curve indicates the outer/middle/inner disc region.  
Note that in the disc-dominated regime  (blu segments) the binary residence time is $\sim 10^9$ yrs, while 
it decreases below $\sim 10^7$ yrs for all binaries, i.e. independent of their mass, at the entrance in the stable region of a circum-binary disc (red dots).
Courtesy of \cite{haiman09}. }
\label{fig:migration}
\end{figure}
 
Semi-analytical expressions of the migration time have been derived in \citep{haiman09} considering orbital decay 
within a Shakura \& Sunyaev accretion disc \citep{shakura73}. This enabled the authors to 
evaluate the disc surface density, opacity, viscosity and ultimately  $M^{\rm edge}_{\rm d}$ as the binary
transits through the outer/middle and inner zones
of the disc. Under these simplifying assumptions (of a steady 1D disc), 
\cite{haiman09} have shown that the sinking time of the binary is a {\it monotonic decreasing function of
the binary orbital period} (or separation). 
The residence time $t_{\rm res}\sim \vert a/{\dot a}\vert$ for equal-mass binaries (which is in this context close to  $\tau^{\rm II}_{\rm mig}$)  is plotted in Figure~\ref{fig:migration} from \cite{haiman09}
for a disc  with  $\alpha$ viscosity parameter equal to 0.3,
a radiative efficiency of 0.1 and an accretion rate equal to 0.1 of the Eddington value. In the disc-dominated regime when
$M_{\rm d}^{\rm edge}>\mbhdue$ the migration timescale is of the order of $\sim$ Gyr and when $M_{\rm d}^{\rm edge}<\mbhdue$ 
it drops below $10^7$ yrs, showing weak dependence of the binary mass.  
Similar timescales have also been found in \cite{rafikov13}
when considering 1D disc models undergoing self-similar evolution [see also figure 6 of \citet{haiman09}].  
Despite these studies,  we are nonetheless far from having
a reliable estimate of the migration timescale in circum-binary discs under a variety of conditions, given the rich physics involved.\footnote{
As an example, in recent studies of planet migration by \cite{duffell14} it has been shown, using highly accurate numerical calculations, that the actual migration rate is dependent on disc and planet parameters, and can be significantly larger or smaller than the viscous drift rate $\tau^{-1}_\nu$. In the
case of disc-dominated migration the rate saturates to a constant value which is in excess of the viscous rate
while in the opposite regime of a low-mass disc, the migration rate decreases linearly with disc mass.}

\begin{figure}
  \includegraphics[width=0.520\textwidth]{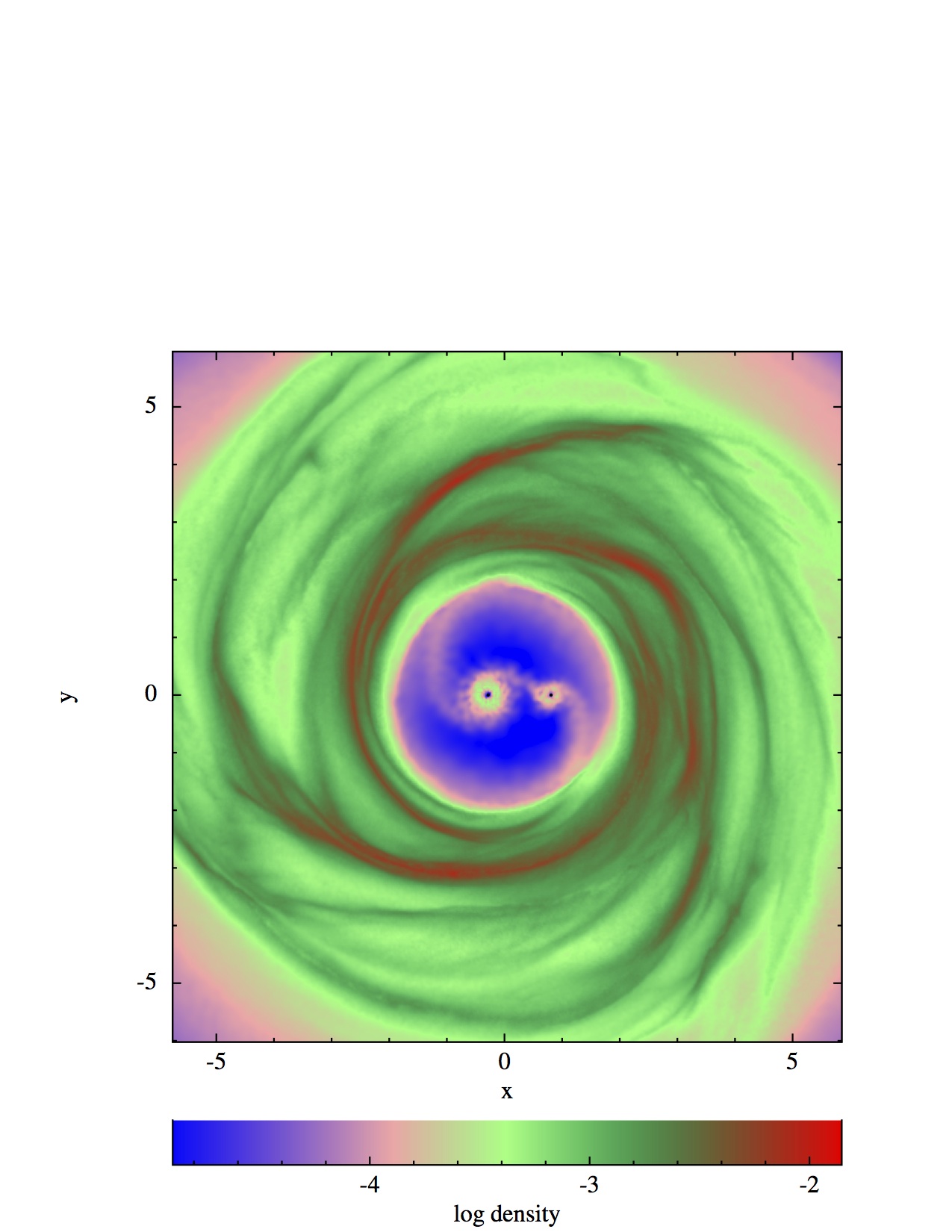}
  \includegraphics[width=0.520\textwidth]{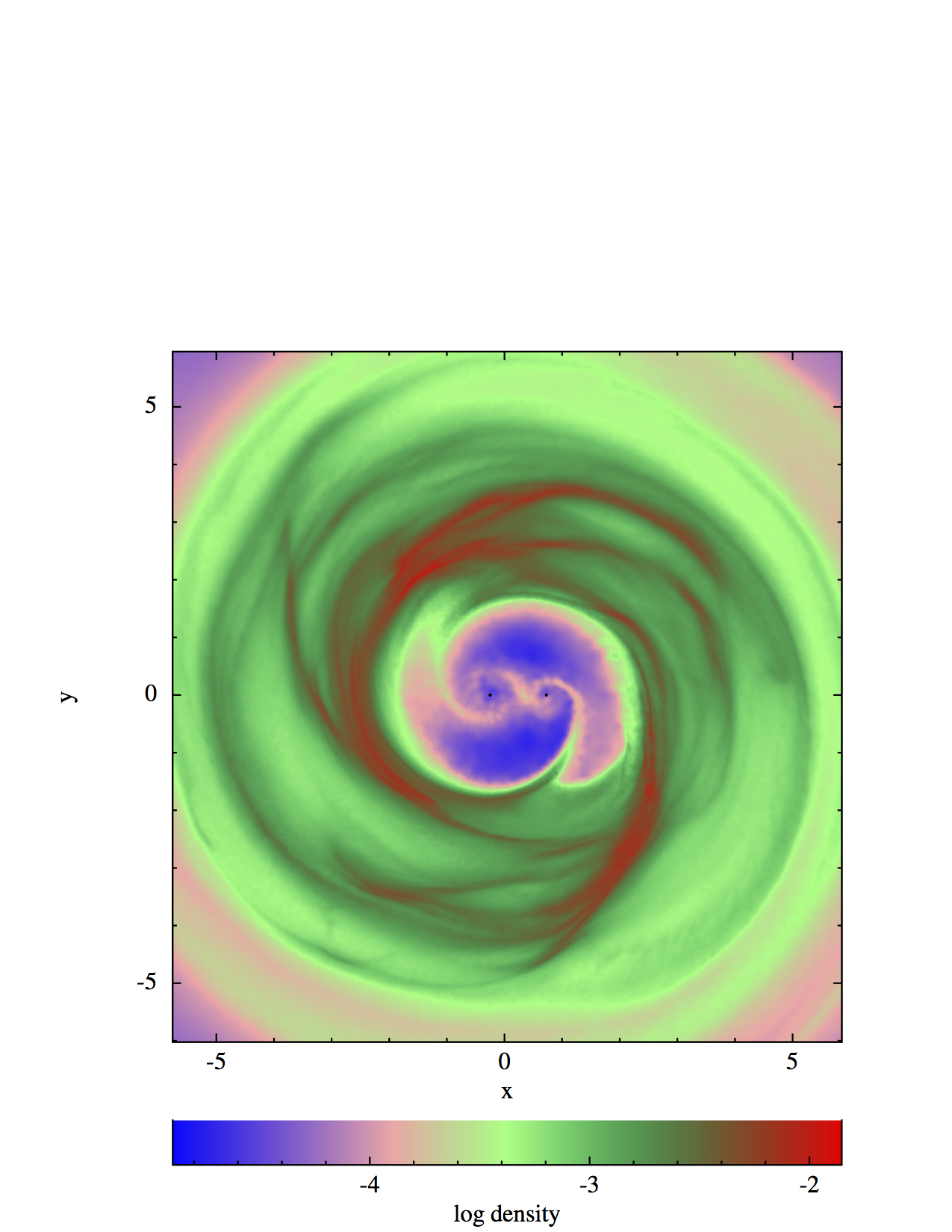}
\caption{Color-coded gas surface density of two Newtonian, self-gravitating circum-binary discs, showing
the presence of a binary region with the two black holes and their mini-discs, a porous cavity filled with streams,
the inner rim or edge working as a dam, and the body disc. Left (right) panel refers to a run with gas in the cavity treated with an
isothermal (adiabatic)  equation of state. Courtesy of \cite{roedig12}. }
\label{fig:gap}
\end{figure}

\begin{figure}
  \includegraphics[width=0.75\textwidth,angle=270]{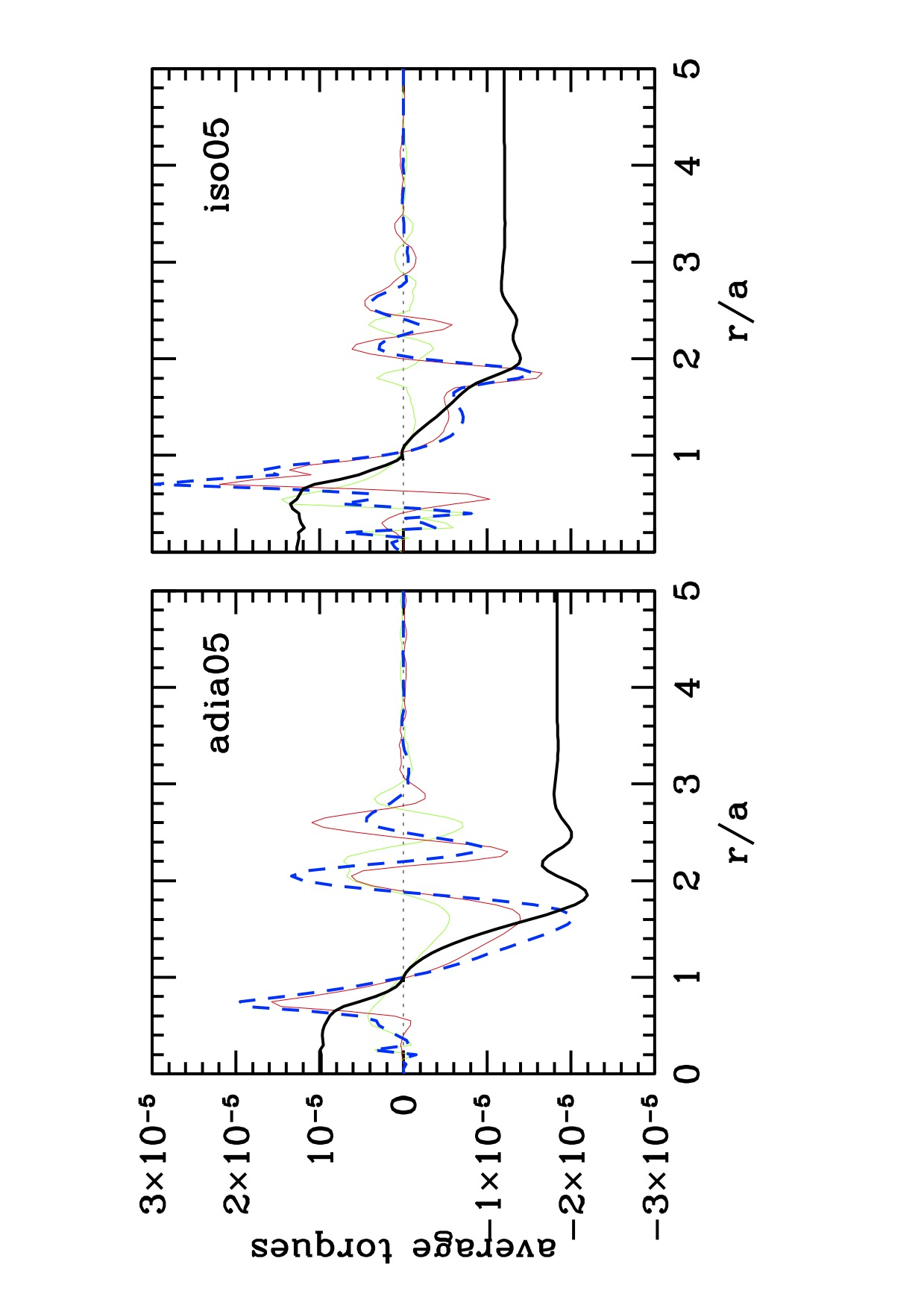}
\caption{Differential torques $dT/dR$ and integrated torque $T$ (averaged over the time span of the simulation, and in code units)
exerted by the disc on the binary with mass ratio $q=0.1$ 
as a function of the radial distance, in units of the binary separation $a$, for the adiabatic (left) and isothermal (right) run from
\citep{roedig12}. In each panel, the differential torque acting on the primary 
is plotted in green, on the secondary in red, and  the sum of the two in blue. Notice that the torque density $dT/dR$ 
shows different signs and starts oscillating around the zero point at distances far from the binary where the 
binary-disc coupling decreases sharply. 
The black line refers to the integrated torque $T$ up to a distance $R$:  $T$ is positive inside $a$, and negative 
outside giving a total negative contribution. Courtesy of \cite{roedig12}. }

\label{fig:torque}
\end{figure}

Gap opening and/or maintenence of the inner cavity around massive black holes
have been seen in numerous numerical simulations of both Keplerian and self-gravitating 
circum-binary discs  \citep{macfadyen08,shikrolik12,cuadra09,roedig11,delvalle12}. 
But interestingly, recent 2D and 3D simulations have 
demonstrated that the binary+disc system  
contains as many as three discs and that these discs may persist being constantly fed by gas flowing through the gap.
The three discs comprise the circum-binary disc plus two mini-discs around each member of the binary \citep{farris14,shikrolik12,roedig12,roedig11}.
This is due to the fact that the disc inner edge is porous (for sufficiently high disc aspect ratios): high velocity, narrow streams of gas leak periodically through the dam into
the inner cavity, modulated by the binary orbit \citep{roedig12,shikrolik12,noble12,orazio13,farris14}.  

Figure~\ref{fig:gap} shows the distribution of gas around the black hole binary after gap formation, from two SPH-3D simulations  of Newtonian, 
massive circum-binary discs \citep{roedig12}  which differ from one anther due to a different thermodynamic modelling of the gaseous streams in the cavity:  isothermal (on the right side) and
adiabatic (on the left side).  The figure highlights the occurrence of different domains in the disc (from outside in): the {\it disc body} $R>2.5a$ where spiral patterns develop; the {\it cavity edge}, at radii $2<R<2.5a,$ which is porous and leaky; the {\it cavity region} or gap, between $a<R<2a,$
which is almost devoid of gas except for the presence of tenuous streams;  the {\it binary} region, at $0<R<a$, with the 
 two black holes and their mini-discs, fed by gas from the disc body flowing through the cavity across the porous dam. 
The mini-discs and the cavity are sharper in the isothermal case compared to the adiabatic case where the amount of gas impacting the gap
is larger. Only a fraction of this gas is captured by the black holes to form the mini-discs, the remaining being swiftly ejected away.
The different  regions highlighted in Figure~\ref{fig:gap} contribute to the differential torque $dT/dR$ on the binary with different signs as illustrated in Figure~\ref{fig:torque}, for the case of a Newtonian
self-gravitating disc (in the adiabatic and isothermal model, respectively).
One can notice that the differential torque shows an oscillatory behaviour with a sharp maximum at the location of the
secondary black hole ($R\sim  0.75a$), and a deep minimum in the cavity region.  Positive and negative peaks alternate in the disc body 
that almost cancel out, giving a negligible contribution to the total torque. Torques on the secondary black hole are always larger than on the primary, due to its proximity to the inner rim of the disc, resulting in a stronger interaction. 

In summary, simulations now indicate that clearing a cavity in the disc does not prevent the inflow of gas through streams across the cavity's edge.  Thus accretion of a fraction of this gas on the black holes, and preferentially onto the secondary
(nearer to the disc's edge) may be a persistent feature \citep{farris14}. 
Thus binary evolution becomes more complex than outlined in the first part of this section. All binary elements 
evolve over time and in some cases inward migration can turn into outward migration \citep{hayasaki09triple,roedig12}. 
The evolution equation for the semi-major axis of a binary depends on changes in the eccentricity, mass, reduced mass and on the 
exchange of angular momentum between the binary and the three discs through a generalised $T$. All these contribute to 
${\dot a}/ a={2T/J_\bullet} -{\dot \mt}/\mt-2{\dot \mu}/ \mu+2e{\dot e}/ (1-e^2)$
where $J_\bullet$ is the binary angular momentum. The sign of this derivative thus depends on different effects.

The binary eccentricity tends to increase during the binary-disc 
coupling  \citep{armitage05ecc,macfadyen08}, and the growth of $e$ has also been seen in  3D numerical simulations \citep{roedig11}.
Progress in the analysis of this process has revealed that such excitation can not grow indefinitely, as {\it saturation} occurs due to
the interaction of the secondary black hole with gas near the inner rim of the disc body, and to
the accumulation of gas around the black holes in the mini-discs \citep{roedig11}.
The initial rise of $e$ can be understood, as in Section~\ref{sec:smoothmigration}, using dynamical friction
in a differentially rotating background as leading argument.
The secondary black hole, closer to the circum-binary disc,  induces
a trailing density wave near the inner rim which reduce its tangential velocity, causing a loss
of orbital angular momentum.  The eccentricity $e$ grows and continues to grow as long as the gas at the inner edge of the circum-binary disc moves with a lower angular velocity. However the progressive decay of the black hole tangential velocity  with increasing $e$ leads eventually to a reversal
of the sign of the relative velocities, the gas moving faster than the black holes,  thus developing a wake heading in front  which leads to an acceleration of the black hole.  The process reaches saturation, 
and this is found to occur about $e\sim 0.6-0.8$. Furthermore, when the binary 
becomes very eccentric, the secondary, less massive black hole passes through the mini-disc of the primary suffering a deceleration
at peri-centre which in turn decreases $e$, which then attains a saturation value.

The mass of the two black holes tend to increase as well, and the increase of the mass of the secondary black hole is even higher,
being closer to the disc, thus driving $q$ toward unity \citep{farris14}. The mass accretion rate is not severely limited (compared to the case of a single isolated black hole)  and $\dot M$ is  found to be  modulated at the binary orbital period and higher harmonics
\citep{farris14,roedig11}.  Interestingly, modulated accretion suggests a promising avenue for producing a modulated
electromagnetic signal permitting the identification of binaries during migration 
in circum-binary discs at different orbital phases along the path to coalescence \citep{eracleous11,decarli13,montuori12}. 

\section{Timescales: an overview}

Galaxy interactions and mergers are the sites of formation of dual, binary, coalescing and
recoiling black holes. 
Associated to these different dynamical phases there is a zoo of sources: the dual, binary and recoiling 
AGN if the black holes are active.   A residence time is associated to each phase:  in phase I, the dynamical
friction timescale $\tau_{\rm df}$ or /and the dynamical friction timescale which accounts for
tidal mass loss $\tau_{\rm df,tidal}$; in phase II,  the hardening time in a stellar background $ \tau_{\rm Hard}$  (which falls in the interval between $\tau^*_{\rm hard}$ and $\tau_{\rm rel}$), or/and (in gas-rich mergers) the timescale of nuclear-disc-driven migration $\tau^{\rm I}_{\rm mig,Mestel}$ and binary-disc-driven
migration $ \tau^{\rm II}_{\rm mig}$; ultimately in phase III, the gravitational wave timescale $\tau_{\rm gw}$.

There is no simple recipe to calculate the residence times in terms of fundamental parameters such as the black hole
mass and mass ratio since these timescales depend on the morphology of
the interacting galaxies, the geometry of the encounter, the gas fraction, and most importantly on the complex input physics of difficult implementation even in current state-of-the-art simulations.  

The characteristic  coalescence time $\tau_{\rm coal}$ would be the sum of the timescales associated to the different phases 
(I, II or I-g,II-g, and III), 
calculated along each individual pathway.  Their value depends, even in the minimal model,
on whether the merger is gas-poor (dry) or  gas-rich (wet), and major or minor. 
As no unique pathway exists for a pair, 
$\tau_{\rm coal}$ can be estimated simply considering the 
maximum of all residence times.  This timescale should then be compared with
the Hubble time or better with the running age of the universe at the time of coalescence, given that eLISA sources are
typically at high redshifts \citep{gwnotes13}.

Here is a tentative summary of the timescales inferred  from the whole body of works, 
in the black hole mass range $\simless 10^7\,\msun$ and for values of the initial black hole mass ratio $q$ (
which indicates the mass ratio between the two interacting galaxies). 
If the "zero" time is calculated when the merger of the baryonic components (bulge and disc) is completed, i.e. when
the black holes
behave as individual objects moving in the relic galaxy,  the relevant timescales in different environments and conditions are expected 
to cluster approximately around these values:

\medskip
\noindent
$\bullet$ In dry major mergers  ($q>q_{\rm crit}\sim 0.1$): (i) dynamical friction time $\tau_{\rm df}\simless \,[10$ Myr - $100$ Myr]\citep{yu02} --
(ii) hardening timescale $\tau_{\rm Hard}\sim 1$ Gyr to a few Gyr \citep{khanjustmerritt11}. 
 
\medskip
\noindent
$\bullet$ In wet major mergers  ($q>q_{\rm crit}\sim 0.1$): (i) gas-dynamical friction time $\tau_{\rm df}\simless \,[10$ Myr - $100$ Myr]\citep{mayer07,chapon13,roskar14} -- (ii) nuclear-disc-driven migration time $\tau_{\rm mig, Mestel}^{\rm I}\sim \,(5$ Myr - $50$ Myr) \citep{escala05,dotti06,dotti07,dotti09,fiacconi13} -- (iii) binary-disc-driven migration
$\tau^{\rm II}_{\rm mig}\sim  10$ Myr \citep{haiman09}. 

\medskip
\noindent
$\bullet$ In dry minor mergers ($q<q_{\rm crit}\sim 0.1$): (i) dynamical friction time $\tau_{\rm df,tidal}\simless \,[10$ Myr - $100$ Myr]\citep{yu02} --
(ii) hardening timescale $\tau_{\rm Hard}\sim 1$ Gyr up to a few Gyr \citep{yu02,khanminor12}.

\medskip
\noindent
$\bullet$ In wet minor mergers ($q<q_{\rm crit}\sim 0.1$): (i) dynamical friction time with
corrections due to tidal stripping $\tau_{\rm df,tidal}\simless \,100$ Myr or wandering \citep{callegari09,callegari11}. 
The fate is uncertain. Depending on the geometry of the encounter and gas fraction, the secondary black hole may wander in the primary galaxy.

\section{Summary and future prospects}

The study of the dynamics of black holes, with masses from $10^4\,\msun$ up to $10^9\,\msun$, inside 
galaxies displaying a large variety of morphologies and masses, is not a side problem:  it is central if we want 
to search for or recognise signs of their duality and/or coalescence at electromagnetic level, and if we want to detect the gravitational waves emitted at 
the time of black hole coalescence. Observationally the search of dual AGN (accreting black holes
in merging galaxies at separations of $\sim \rm kpc$), and of binary (pc scales) and recoiling 
AGN 
have received attention in recent years 
\citep{eracleous11,liu11,komossa12,koss12,liu13,decarli13,comerford14,lusso14,fukun14}.
There has been some major advances in the study of the dynamics of black holes in merging galaxies, over the last years, and 
the points to remember and to take away for future reference are:

\begin{enumerate}

\item{Black holes in binaries can reach coalescence under the emission
of gravitational waves.  But, for this to happen, the black holes have to be driven to 
separations as small as $\sim 10^{-3}$ pc or less, as gravity is a weak force and gravitational waves are
a manifestation of the strong field regime \citep{sathya09}.
This is a minuscule distance, compared to galaxy's sizes, and merging galaxies are the
sites where these events can occur. Nature has thus to provide a series of mechanisms
able to extract energy and angular momentum, from the large scale of the merger (at least a few kpc) to
the micro-parsec scale, i.e. the scale at which the black hole horizons touch.
The path to coalescence is long and complex, and stalling of the binary at some scale is a possibility. }

\item{Three phases accompany the path to coalescence: the pairing, hardening, and gravitational-wave driven inspiral phases.
Stars or gas, or stars and gas drive the black hole inspiral, depending on whether galaxies are  gas-rich or gas-poor. Bottlenecks can appear at various scales and a major effort is to identify possible obstacles.  The last parsec problem, i.e.
the stalling of a massive black hole binary at the centre of a large spherical collisionless galaxy was highlighted as a critical step. }

\item{Thanks to recent advances in numerical computing, the last parsec problem 
appears to be an artefact of oversimplifying assumptions.  Galaxies, relic of mergers, are not spherical systems
 and can retain a high degree of triaxiality or asymmetry. Under these circumstances the hardening of the binary via binary-single stellar encounters appears to have no halt, at least in the cases explored and coalescence timescales are close to 1 to a few Gyr. The issue is not settled yet.}

\item{ It is now possible to track the black hole
dynamics during galaxy collisions using state-of-the-art simulations. 
The dynamics of the 
interacting galaxies is followed {\it ab initio}, 
from the large scale (several kpc)  down to the central few parsecs, considering 
all components - the dark halo, the stellar and gaseous disc, and bulge.
This enables us to trace the rise of asymmetries and instabilities 
in both the stellar and gas components which 
play a pivotal role in determining whether there is stalling or rapid sinking of the binary.}

\item{Major gas-rich (wet) mergers are conducive to the formation of close Keplerian binaries. The gas, thanks to its 
high degree of dissipation, controls the black hole inspiral inside the 
massive circum-nuclear disc that forms in the end-galaxy. When described as a single phase medium, the gas promotes rapid inspiral before
the disc fragments into stars, and before stellar dynamical friction becomes effective. 
When the gas is multi-phase and clumpy on various mass scales, 
the black hole orbit shows a stochastic behaviour. The black holes in this case form a binary on timescales typically between 1 Myr and 100 Myr.}

\item{Coalescences of middleweight black holes of $\sim 10^4\,\msun$ at high redshifts $z\sim 10-15$ require either very dense,
low velocity dispersion stellar
environments, or large not yet quantified amounts of gas in unstable forming galaxies.}

\item{Minor mergers can release the less massive black hole on peripheral orbits in the main galaxy, due to
the disruptive action of tidal torques on the less massive galaxy, rising a new problem:  the last kpc problem. 
Gas plays a key role in the process of pairing in minor merger, as it makes the satellite galaxy more resilient
against tidal stripping because central gas inflows, triggered during the interaction, steepen the stellar cusp. 
Due to the fragility of the satellite galaxy, the fate of black holes in minor mergers is uncertain: encounter geometry, gas fraction, degree of gas dissipation are key elements for establishing whether
the black hole is a {\it sinking} or a {\it wandering} black hole inside the primary galaxy. The prediction is that there is a large scatter in the outcomes.}

\item{At sub-parsec scales, the Keplerian binary is likely surrounded by a circum-binary disc. When present, gas-assisted inspiral takes place which 
can be faster than star-driven inspiral. Both processes likely co-exist but have never been treated jointly. 
The black holes are expected to migrate on a timescale controlled by the interplay between the binary tidal torque which tends to 
clear a cavity (repelling the disc's gas in the immediate vicinity of the binary) and 
viscous torques in the disc which tend to fill and even overfill the cavity.
Gas leaks through the gap and the black holes are surrounded by mini accretion discs. All these processes modify the orbital elements in a complex way. 
Theoretical models indicate that  if there is a sufficiently long-lived inflow of gas at the inner edge of the circum-binary disc, the binary
hardens on timescales $\simless 10^9$ Gyr or even much less (depending on the previous history).}

\item{The path to coalescence still remains a complex problem to solve and there is no clear-cut answer.  
To be conservative, coalescence times range between several Myrs to about several Gyrs.}

\end{enumerate}

\medskip
\noindent
The field needs to evolve farther along different lines and directions. Here are a few hints:

\begin{enumerate}
\item{There is need to continue to study not only the growth of black holes along cosmic history,
but their dynamics as they are inter-connected. Attempts to follow
the dynamics of black holes during the cosmic assembly of galactic halos have been carried on, albeit at much lower
revolution than required to track their accretion history and fate \citep{bellovary10}.  Any effort along this line is central in order to
understand the coevolution of black holes with galaxies.}

\item{Intriguingly enough, the fate of black hole binaries in galaxies takes us back to the 
unsolved problem of the feeding of black holes in galactic nuclei over cosmic ages, i.e. of the angular momentum barrier present on 
parsec scales.  Dynamical decay, star formation, accretion and their back-reactions are coupled.
Star formation makes
the ISM multi-phase and turbulent. Supernova and AGN feed-back may heat/remove gas and the consequences of these effects
on black hole migration have not been quantified yet, during orbital evolution.}

\item{Dual, binary and recoiling AGN and also triple AGN are important observational targets. There is the need to improve 
upon observational strategies for identifying binary and recoiling AGN in large surveys, assisted by tailored and coordinated 
hydro-dynamical simulations.}

\item{Black hole migration in circum-binary disc is a challenging problem which deserves constant attention. Binary eccentricity growth, accretion and outflows, are processes that affect the dynamics and stability of the system as a whole.  Understanding the nature of torques 
in multi-phase models of circum-nuclear discs will become central in order to extrapolate the black hole migration timescale 
from the parsec scale down to the domain controlled by gravitational wave inspiral, and to asses the observability of
sub-parsec binaries during the phases which anticipate the merging.}

\end{enumerate}

\bigskip
\noindent
{\bf Acknowledgements} I would like to thank my collaborators Simone Callegari, Massimo Dotti, Davide Fiacconi, Alessandro Lupi, Lucio Mayer, Constanze Roedig, Alberto Sesana 
and Marta Volonteri for many useful and illuminating discussions over the years. I would like also to thank the International Space Science Institute for kind hospitality.  
\bibliographystyle{plainnat}
\bibliography{Biblio}

\end{document}